\documentclass[referee,a4paper,12pt,traditabstract]{jswsc} 

\usepackage{graphicx}
\usepackage{txfonts}
\usepackage{subfigure}
\usepackage{epstopdf}
\usepackage[displaymath,mathlines]{lineno}
\usepackage[authoryear,round]{natbib}
\usepackage[backref]{hyperref}
\usepackage{url}
\usepackage{multirow}
\usepackage[table]{xcolor}
\usepackage{longtable}
\usepackage{tabularx}
\usepackage{placeins}
\usepackage{float} 
\usepackage{caption}

\hypersetup{colorlinks=true,citecolor=cyan,urlcolor=cyan,linkcolor=blue}

\definecolor{lightgray}{gray}{0.9}

\begin{document}


   \title{Using sunRunner3D to interpret the global structure of the heliosphere from in-situ measurements}
   
   \titlerunning{sunRunner3D simulations}
   
   \authorrunning{Gonz\'alez-Avil\'es et al.}

   \author{J. J. Gonz\'alez-Avil\'es \inst{1}
          \and
          P. Riley \inst{2}
          \and
          M. Ben-Nun \inst{2}
          \and
          P. Mayank \inst{3}
          \and
          B. Vaidya \inst{3}
          }

\institute{Escuela Nacional de Estudios Superiores, Unidad Morelia, Universidad Nacional Aut\'onoma de M\'exico, 58190 Morelia, Michoac\'an, M\'exico
             \email{\href{jgonzaleza@enesmorelia.unam.mx}{jgonzaleza@enesmorelia.unam.mx}}
             \and
             Predictive Science Inc., San Diego, CA 92121, USA\\
             \email{\href{pete@predsci.com}{pete@predsci.com}}
             \and
             Department of Astronomy, Astrophysics and Space Engineering, Indian Institute of Technology Indore, India\\
             \email{\href{phd2001121004@iiti.ac.in}{phd2001121004@iiti.ac.in}}
             }

 
  \abstract   
 {Understanding the large-scale three-dimensional structure of the inner heliosphere, while important in its own right, is crucial for space weather applications, such as forecasting the time of arrival and propagation of coronal mass ejections (CMEs). This study uses sunRunner3D (3D), a 3-D magnetohydrodynamic (MHD) model, to simulate solar wind (SW) streams and generate background states. SR3D employs the boundary conditions generated by CORona-HELiosphere (CORHEL) and the PLUTO code to compute the plasma properties of the SW with the MHD approximation up to 1.1 AU in the inner heliosphere. We demonstrate that SR3D reproduces global features of Corotating Interaction Regions (CIRs) observed by Earth-based spacecraft (OMNI) and the Solar TErrestial RElations Observatory (STEREO)-A for a set of Carrington rotations (CRs) that cover a period that lays in the late declining phase of solar cycle 24. Additionally, we demonstrate that the model solutions are valid in the corotating and inertial frames of references.
 Moreover, a comparison between SR3D simulations and in-situ measurements shows reasonable agreement with the observations, and our results are comparable to those achieved by Predictive Science Inc.'s Magnetohydrodynamic Algorithm outside a Sphere (MAS) code. We have also undertaken a comparative analysis with the Space Weather Adaptive Simulation Framework for Solar Wind (SWASTi-SW), a PLUTO physics-based model, to evaluate the precision of various initial boundary conditions. Finally, we discuss the disparities in the solutions derived from inertial and rotating frames.}
 


   \maketitle

\section{Introduction}
\label{Introduction}

Knowledge about the physical properties of the plasma from the solar atmosphere is crucial for understanding the solar wind (SW) dynamics and explosive events, such as coronal mass ejections (CMEs). These phenomena connect the Sun and the Earth, leading to an essential area of research known as Space Weather. For example, the CMEs cause geomagnetic storms that could affect the Earth's magnetic field environment. These geomagnetic storms can disrupt the electric power supply, perturb navigation systems, and interrupt satellite functionality on Earth. Moreover, these events could impact the global economy, making it essential to develop novel scientific research dedicated to space weather forecasting \citep{Schrijveretal2015}.  

In-situ observations of the core properties of the SW and CMEs provide potentially useful scientific and operational information. In particular, e.g., the Solar and Heliospheric Observatory (SOHO), the Solar TErrestial RElations Observatory (STEREO), the Solar Dynamic Observatory (SDO), and the Global Oscillation Network Group (GONG) produce valuable information about the solar surface. In addition, missions such as the Advanced Composition Explorer (ACE), WIND, and the Deep Space Climate Observatory (DSCOVR) cover the near-Earth region. Finally, the Parker Solar Probe (PSP) and Solar Orbiter (SolO) observe the properties of the SW and CMEs in the inner heliosphere. These observatories provide in-situ measurements of physical properties, such as velocity, proton density, temperature, and magnetic field of the SW and CMEs. However, they do not accurately estimate key forecasting parameters, such as arrival times of SW currents and CMEs near the Earth, i.e., at about 1 AU. Therefore, many current investigations employ numerical models to improve these limitations to develop a broader picture of space weather events, such as SW streams, CMEs, solar energetic particles (SEPs), Corotating Interaction Regions (CIRs), and stream interaction regions (SIRs). 

There is a noticeable advance in models applied to study the propagation and dynamics of the SW and the interplanetary counterpart of the CMEs, commonly called Interplanetary Coronal Mass Ejections (ICMEs), in the inner heliosphere. Generally, we classify these models into three main categories: empirical, semi-empirical, and numerical. Empirical models use a probabilistic forecasting approach to analyze SW observations at Sun-Earth L1 Lagrangian point \citep[see, e.g.,][]{BussyRidley2014, Rileyetal2107,2017SoPh..292...69O}. Semi-empirical models employ semi-empirical relations of SW speed based on observations of coronal holes \citep[see, e.g.,][]{https://doi.org/10.1002/2016SW001390,https://doi.org/10.1029/1999JA000262}. Numerical models are based on photospheric magnetograms to determine SW plasma properties in the inner heliosphere. These models numerically solve the MHD equations to describe the propagation and evolution of the SW streams and CMEs in the inner heliosphere, and among these models are, for example, MAS \citep{Riley_et_al_2001}, ENLIL \citep{2003AdSpR..32..497O}, Space Weather Modeling Framework \citep[SWWF,][]{Toth_et_al_2005}, Solar–Interplanetary space–time conservation element and solution element \citep[SIP-CESE,][]{Feng_et_al_2010}, Space-weather-forecast-Usable System Anchored by Numerical Operations
and Observations \citep[SUSANOO,][]{Shiota_et_al_2014}, EUropean Heliospheric FORecasting Information Asset \citep[EUHFORIA,][]{Pomoell&Poedts_2018}, and more recently, SWASTi-SW \citep{Mayank_2022}. However, despite all these models, there is still no model capable of capturing the fundamental parameters such as velocity, density, temperature, magnetic field polarity, and magnetic field strength of SW that fully match with in-situ observations. Hence, there are still opportunities to continue exploring new models

In this paper, we use sunRunner3D (SR3D) to interpret the global structure of the inner heliosphere in terms of SW from in-situ measurements. SR3D is a community-developed open-source package and an improvement of sunRunner1D \citep{Riley&Ben-Nun_2022}, which is a tool for exploring ICMEs evolution through the inner heliosphere, considering spherical symmetry in 1D. We plan to apply SR3D to specific scientific problems in heliophysics research. For example, to interpret the signatures of CMEs observed by various heliospheric spacecraft, including PSP and SolO. It can also be applied to explore the dynamic evolution of CME events out to 1 AU and to investigate whether CME events would have produced extreme space weather phenomena if they were directed toward the Earth. Besides, recently, in \cite{Aguilar-Rodriguez_et_al_2024}, SR3D has been successfully used to perform numerical global MHD simulations of SIRs and CIRS observed by PSP and STEREO-A.

In particular, for this paper, SR3D uses the boundary conditions generated by CORHEL \citep{Riley_et_al_2012, Linker_2016} to drive the heliospheric model. In the inner heliosphere domain, that goes from 0.14 AU to 1.1 AU, SR3D employs the PLUTO code \citep{Mignone_et_al_2007} to compute the plasma properties of SW with the MHD approximation in the inertial and rotating frames of reference. One of the primary purposes of this paper is to simulate SW streams and produce background states in spherical coordinates in three dimensions in both inertial and rotating frames and compare with in-situ measurements observed by Earth-based spacecraft (OMNI dataset) and STEREO-A (STA), and with the MAS code for a set of Carrington rotations that cover a period that lays in the declining phase of solar cycle 24.

We organize the paper as follows. First, in Section \ref{Model}, we describe the CORHEL and PLUTO models. Then, in Section \ref{Results}, we present the results of the numerical simulations of the relaxed, steady-state SW solutions, the comparisons of the model results with in-situ measurements of OMNI and STA, the statistical analysis for a set of Carrington Rotations, and the comparison with similar existing model. Finally, we draw our conclusions in Section \ref{Conclusions}.

\section{Model}
\label{Model}

SR3D is an MHD model to calculate SW stream steady-state solutions and evolving CMEs in a 3D spherical coordinate system in the inner heliosphere in the inertial and rotating frames of reference. Specifically, it contains two main parts: i) boundary conditions and ii) heliospheric model. The CORHEL model provides the boundary conditions described below, while the PLUTO code constitutes the MHD heliospheric model. The medium and long-term plans are to distribute SR3D as a community-developed open-source package.

\subsection{The CORHEL model}
\label{Corhel_mas_model}

To generate the boundary conditions, we used the CORHEL framework, which is capable of modeling the ambient solar corona and the inner heliosphere for a specific period of interest. Mainly, it derives the boundary conditions using maps of the Sun's photospheric magnetic field derived from magnetograms. These magnetograms are obtained principally from the Helioseismic and Magnetic Imager (HMI) instrument of the SDO. Then, it runs the coronal model using the MAS code \citep{1996AIPC..382..104M,1999PhPl....6.2217M} until obtaining a relaxed state, which serves to produce the boundary conditions for the heliospheric models. CORHEL solutions are available to the community at the Community Coordinated Modelling Center (CMMC, \href{https://ccmc.gsfc.nasa.gov}{https://ccmc.gsfc.nasa.gov}) and the Predictive Science website (\href{https://www.predsci.com}{https://www.predsci.com}). 

In this paper, we take advantage of the fact that boundary conditions can be downloaded directly from \href{https://www.predsci.com/data/runs/}{https://www.predsci.com/data/runs/}. There, the boundary conditions are already in a readable format for the PLUTO code, so we have all the variables, such as number density, radial velocity, radial magnetic field, and temperature, to set at the inner boundary ($R_{b}=0.14$ AU) and the whole domain as initial conditions. We use the relaxed steady-state solution of the variables to drive the inner heliospheric MHD model, described in the following subsection. 

This paper uses the MAS polytropic solutions obtained by solving the set of MHD equations in spherical coordinates, employing an adiabatic energy equation. To reproduce the structure of the magnetic field and generate solutions with sufficient variation in SW speed or densities, the model employs the empirically based approach called the "Distance from the Coronal Hole Boundary" (DCHB), which helps to specify the SW speed at the inner boundary of the chosen heliospheric code \citep{Riley_et_al_2001}. The DCHB model assumes the flow is fast within coronal holes (i.e., away from the boundary between open and closed magnetic field lines). The flow is slow at the boundary between open and closed field lines. Over a relatively short distance, the rise of the flow speed is smoothed to match the fast coronal hole flow \citep[see, e.g.,][]{Riley_et_al_2001}. In summary, the MAS polytropic solutions, which incorporate the empirical DCHB model and are relaxed, serve as boundary conditions for SR3D. These polytropic solutions ensure consistency with the inner heliospheric MHD model, which utilizes the polytropic approximation with $\gamma = 5/3$ to characterize the plasma properties of the solar wind. 

\subsection{The inner heliosphere MHD model}
\label{MHD_model}

The inner heliosphere model employs the PLUTO code from $R_{b}=0.14$ AU outwards to solve the three-dimensional time-dependent MHD equations in spherical coordinates. Particularly, we adopt the ideal MHD equations written in the following dimensionless conservative form,

\begin{eqnarray}
\frac{\partial\varrho}{\partial t} + \nabla\cdot(\varrho{\bf v}) &=& 0, \label{density}\\
\frac{\partial(\varrho{\bf v})}{\partial t} + \nabla\cdot(\varrho{\bf v}{\bf v}-{\bf B}{\bf B} + p_{t}{\bf I}) &=& \varrho{\bf g} + {\bf F},  \label{momentum} \\
\frac{\partial E}{\partial t} +\nabla\cdot((E+p_{t}){\bf v}-{\bf B}({\bf v}\cdot{\bf B})) &=& \varrho{\bf v}\cdot{\bf g} + {\bf v}\cdot{\bf F}, \label{energy} \\
\frac{\partial{\bf B}}{\partial t} +\nabla\cdot({\bf v}{\bf B} -{\bf B}{\bf v}) &=& 0, \label{evolB} \\
\nabla\cdot{\bf B} = 0, \label{divB}
\end{eqnarray}

\noindent where $\varrho$ is the plasma density, ${\bf v}$ represents the fluid velocity, $E$ is the total energy density, ${\bf B}$ is the magnetic field, $p_{t}$ is the total pressure (thermal+magnetic), and ${\bf I}$ is the unit matrix. In the equations, $p_{t}=p+{\bf B}^{2}/2$, where by the ideal gas law $p= \varrho k_{B} T/ {\bar{m}}$. Here $T$ is the temperature of the plasma, $\bar{m}=\mu m_{H}$ is the particle mass specified by a mean molecular weight value $\mu=$0.6 for a fully ionized gas, $m_{H}$ is the mass of the hydrogen atom, and $k_{B}$ is the Boltzmann constant. In addition, $E= p/(\gamma-1) + \varrho{\bf v}^{2}/2 + B^{2}/2$, being $\gamma=5/3$ the polytropic index. The source terms include the gravitational forces defined in terms of the gravitational acceleration \begin{math}\mathbf{g} = -(GM_\odot/r^{2})\mathbf{\hat{r}}\end{math}, with $G$ representing the gravitational constant and $M_{\odot}$ the solar mass. Besides, $\bf F$ contains the Coriolis and centrifugal forces that in PLUTO code are treated conservatively \citep{1998A&A...338L..37K}.

Equations (\ref{density})-(\ref{divB}) are solved in a 3D spherical coordinate system $(r, \theta, \phi)$ for both rotating and inertial frames of reference. In the case of the solution in the rotating frame, we consider ${\bf F}$ and let the system rotate with constant angular velocity $\Omega_{c} = 2.8653\times10^{-6}$ Hz, which represents the rotation rate of the solar equator. In the rotating frame, we define the azimuth component of the velocity $V_{\phi}=-r\Omega_{c}$, in the whole domain and at the boundary $R_{in}=$0.14 AU. This definition helps us better match between solutions with PLUTO in the inertial and rotation frames, as we will see in the simulation results. On the other hand, in the inertial frame, we set ${\bf F=0}$ and account for the solar rotation by rotating the boundary conditions in $\phi$ (longitude) at a rate equal to $\Omega_{c}$, which is similar to the method employed in CORHEL, ENLIL, and EUHFORIA. Finally, to ensure a zero electric field (${\bf v}\times{\bf B} = 0$) in the rotating frame, we introduce an azimuthal component of the magnetic field $B_{\phi} = -B_{r}\sin\theta\left(v_{rot}/v_{r}\right)$, where $v_{rot}=\Omega r$, representing the rotating speed of the inner boundary. The latter definition of $B_{\phi}$ is consistent with the requirement for a steady-state solution in the rotating frame \citep[see, e,g.,][]{Pomoell&Poedts_2018, Mayank_2022}. 

The MHD domain extends from 0.14 to 1.1 AU along the radial coordinate, $0^{\circ}$ to $180^{\circ}$ in latitude, and $0^{\circ}$ to $360^{\circ}$ in longitude, with a grid resolution of $141\times111\times128$, respectively. We chose the RK2 time-stepping algorithm for the time integration, a second-order linear reconstruction scheme in combination with minmod limiter, and the Harten–Lax–van Leer–Contact (HLLC) Riemann solver \citep{Li_2005}. To ensure the divergence-free condition (equation [\ref{divB}]), we selected Powell's eight-wave formulation \citep{Powell1997}. At the inner-radial boundary, we specify the values of the radial velocity $v_{r}$, the number density $n$, the temperature, and the radial component of the magnetic field $B_{r}$, given by the CORHEL coronal solution at $R_{in}=0.14$ AU. At the outer radial boundary, we set outflow boundary conditions, which allow the fluid to leave the computational domain. Furthermore, we employ polar axis boundary conditions at both poles, which works in conjunction with the ring average technique that helps to remove time step restrictions near a singular axis, i.e., at $\theta =0,\pi$ \citep[see, e,g.,][]{ZHANG2019276}. Finally, we set periodic boundary conditions in longitude.

\section{Simulation results}
\label{Results}

\begin{figure*}
\centering
\includegraphics[width=7.0cm,height=6.0cm]{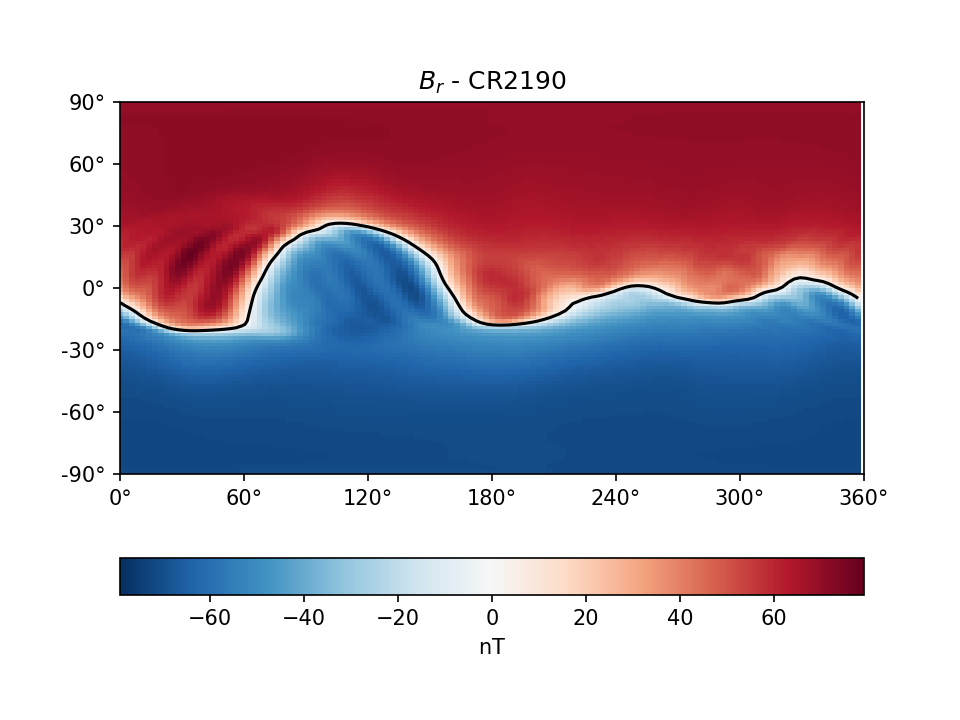}
\includegraphics[width=7.0cm,height=6.0cm]{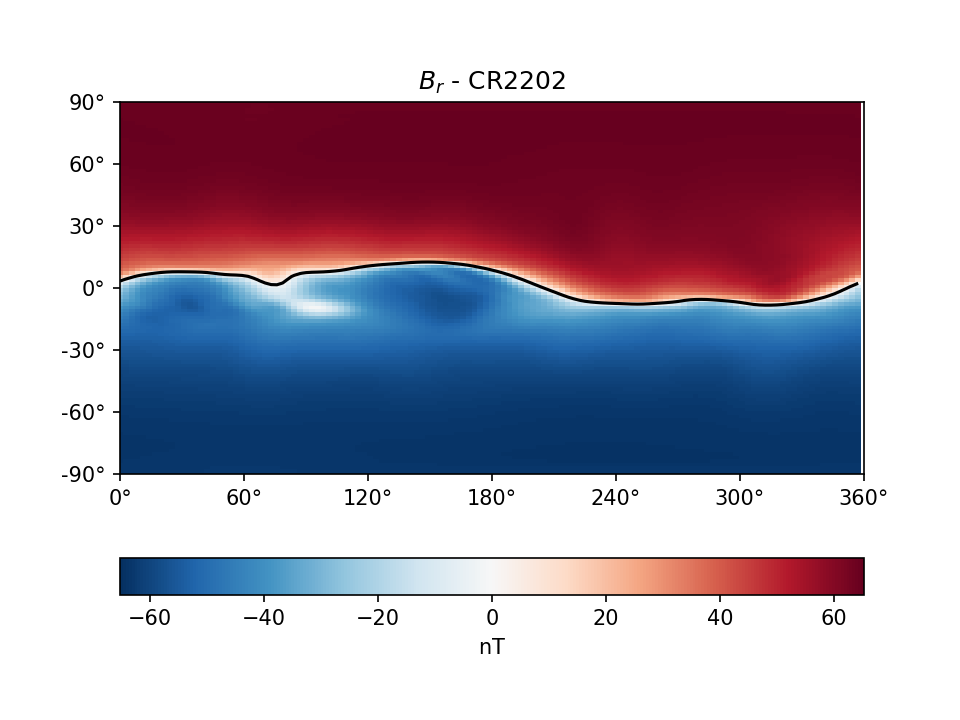}\\
\includegraphics[width=7.0cm,height=6.0cm]{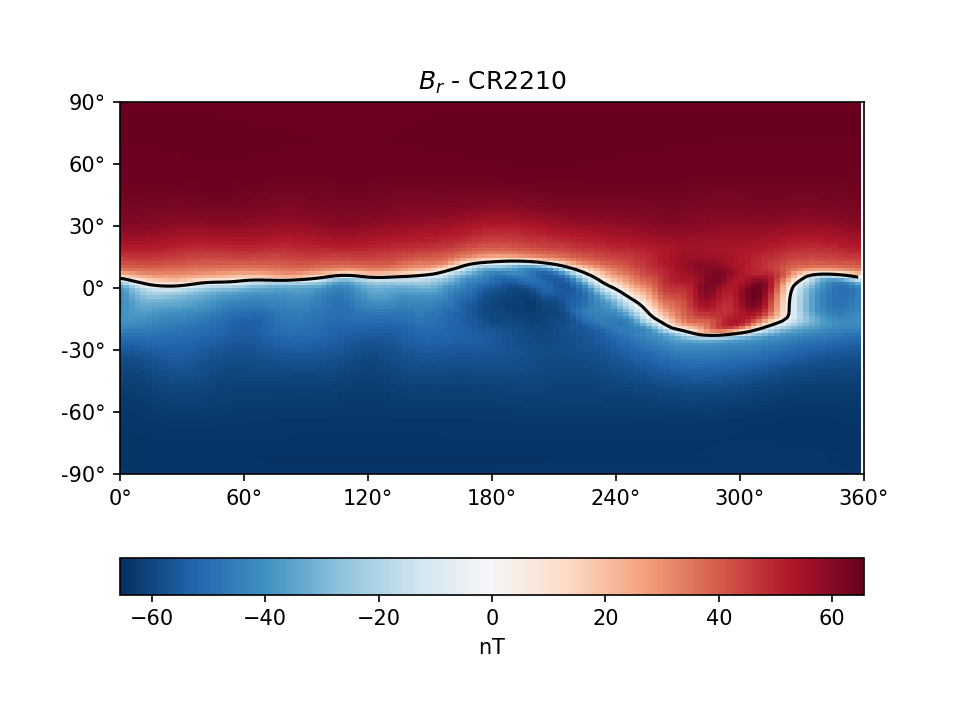}
\includegraphics[width=7.0cm,height=6.0cm]{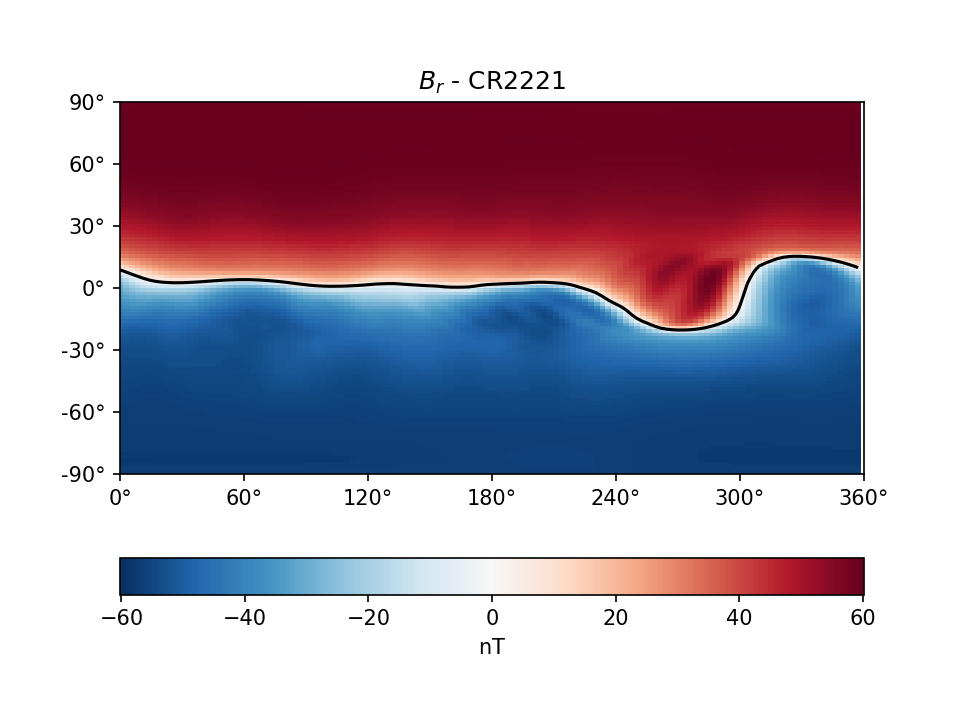}
\caption{Synoptic maps of the radial magnetic field $B_{r}$ (nT) at the inner radial boundary ($\sim$0.14 AU) for CR2190 (top-left), CR2202 (top-right), CR2210 (bottom-left), and CR2221 (bottom-right). For each panel, the horizontal axis is the longitude, and the vertical axis is the latitude in the Carrington coordinate system. The black line in each panel is the contour of the radial magnetic field drawn at $B_{r}=0$, which marks the location of the heliospheric current sheet.}
\label{synoptic_maps}
\end{figure*}

We select the following CRs to validate SR3D: CR2190, CR2199, CR2202, CR2203, CR2205, CR2208, CR2209, CR2210, CR2211, CR2214, CR2215, and CR2221. These CRs are between April 2017 and August 2019, coinciding with the declining phase of cycle 24. In Figure \ref{synoptic_maps}, we show the synoptic maps of $B_{r}$ with a resolution of $180\times360$ at about $r=30.1045 R_{\odot}\sim 0.14$ AU obtained with CORHEL for some representative CRs. These maps are interpolated to a coarser 3D spherical grid of PLUTO code and represent the input of the MHD model at the inner boundary ($r\sim$0.14 AU). Throughout the manuscript, we name the corresponding solutions for PLUTO in the inertial and rotating frames as SR3D-I and SR3D-R, respectively.      

\begin{figure}[h]
\centering
\includegraphics[width=5.8cm,height=5.5cm]{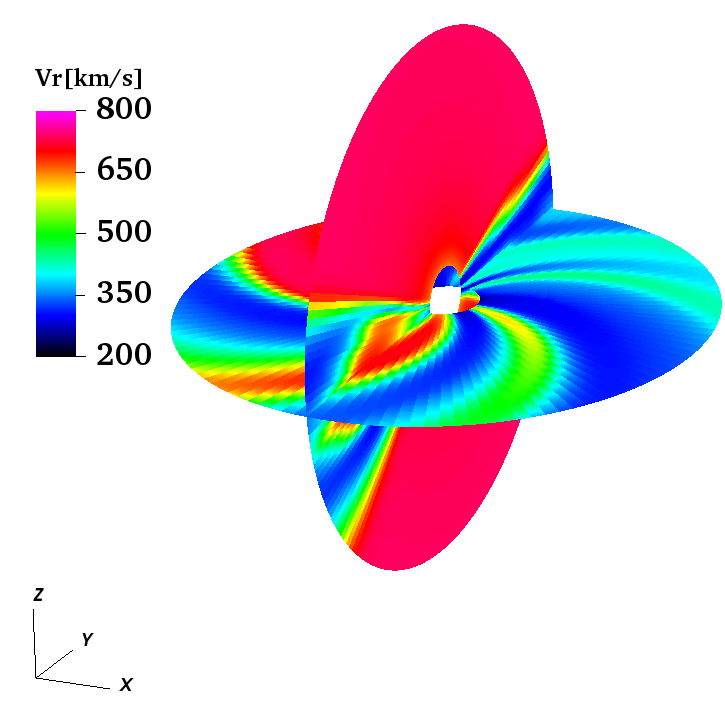}
\includegraphics[width=5.8cm,height=5.5cm]{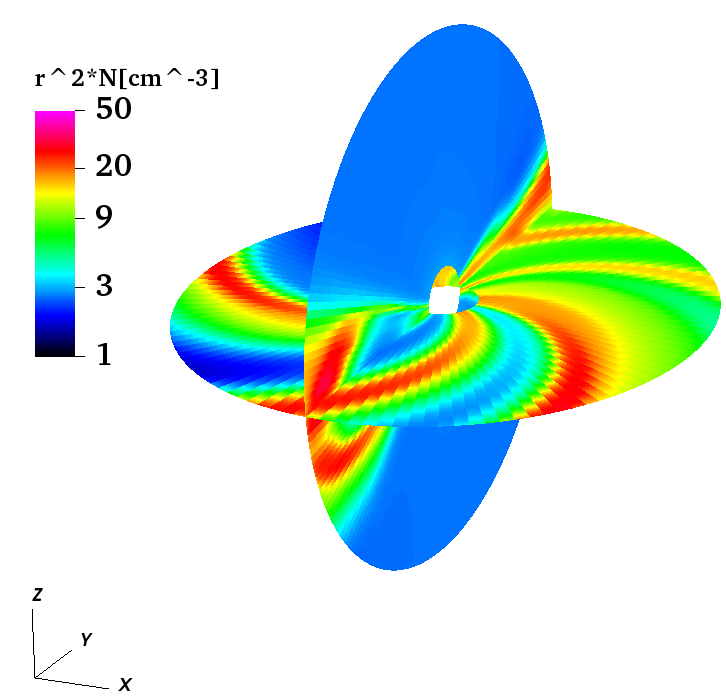}\\
\includegraphics[width=5.8cm,height=5.5cm]{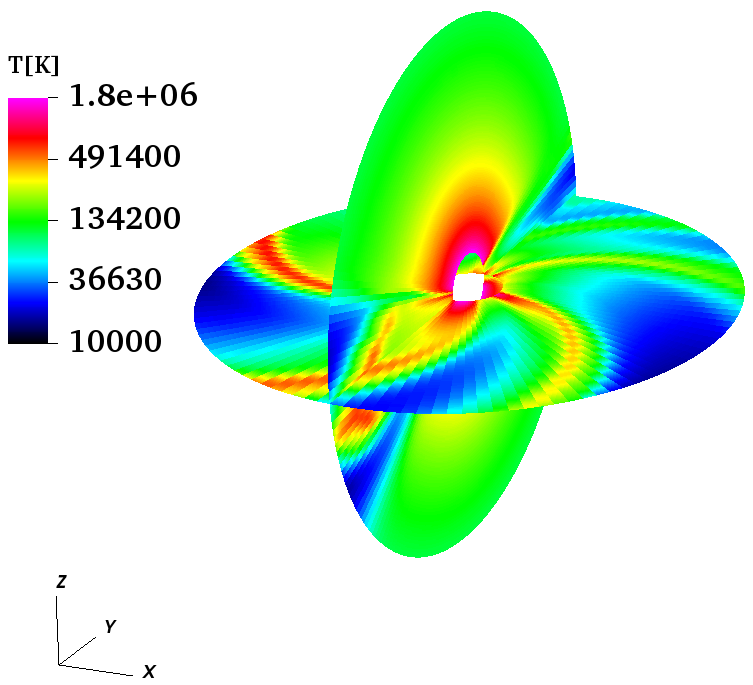}
\includegraphics[width=5.8cm,height=5.5cm]{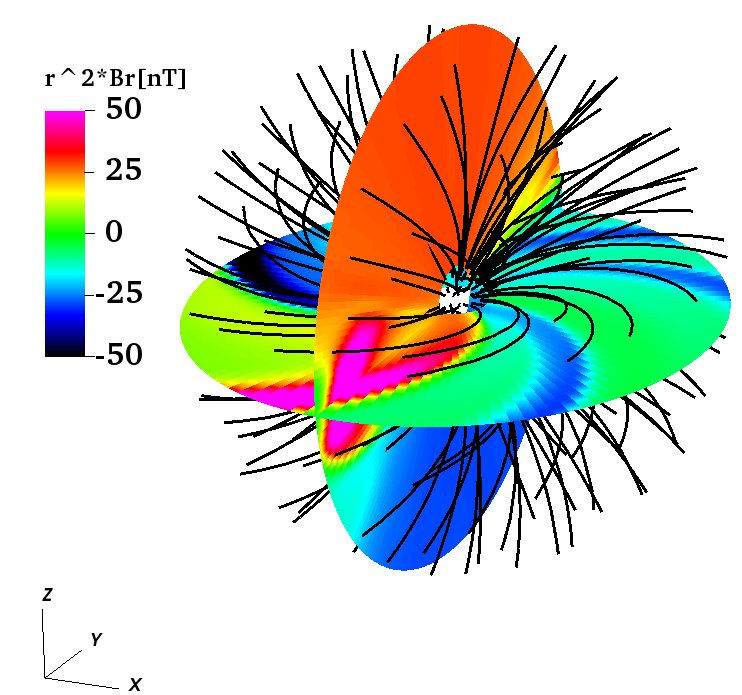}
\caption{Maps of simulated SW parameters in the heliospheric equatorial and meridional planes for CR2210 in the rotating frame. From the upper left to the lower right, the panels show velocity $V_{r}$ (km s$^{-1}$), scaled number density $r^{2}N$ (cm$^{-3}$), temperature $T$ (K), and the scaled radial magnetic field $r^{2}B_{r}$ (nT), overlay with magnetic field lines (black).}
\label{CR2210_3D}
\end{figure}

In Figure \ref{CR2210_3D}, we show the results for the relaxation of the SW conditions corresponding to CR2210 using the solution of SR3D-R. For this simulation, we ran out PLUTO long enough to convect any/all transient phenomena created at $t=0$ past the outer radial boundary. In this figure, we normalize $N$ multiplying by the factor $r^{2}$, which helps to illustrate the results better. In particular, we display equatorial and meridional cuts, observing standard features of interplanetary solutions expected for the steady state SW. We recognize the fast and slow SW streams in the radial velocity, which are discernible as low and high-density SW streams, respectively in the normalized number density maps. Notably, in the plot of the radial velocity $V_{r}$, it is also visible that the high-speed ($>700$ km s$^{-1}$) SW dominates in the north and south poles, as shown in the meridional plane. We also observe a mix of slow and fast wind at all latitudes, which the coronal structure during the period of CR2210 could produce. In addition, the temperature cuts are also a mix of low and high-temperature regions, which are related to the presence of high and low-density SW streams there. Finally, in the radial magnetic field cuts, we identify the formation of the Parker spiral represented by the magnetic field lines colored in black. 

\begin{figure}[h]
\centering
\centerline{\bf   
      \hspace{0.48\textwidth}  \color{black}{(A)}
         \hfill}
\includegraphics[width=0.65\textwidth]{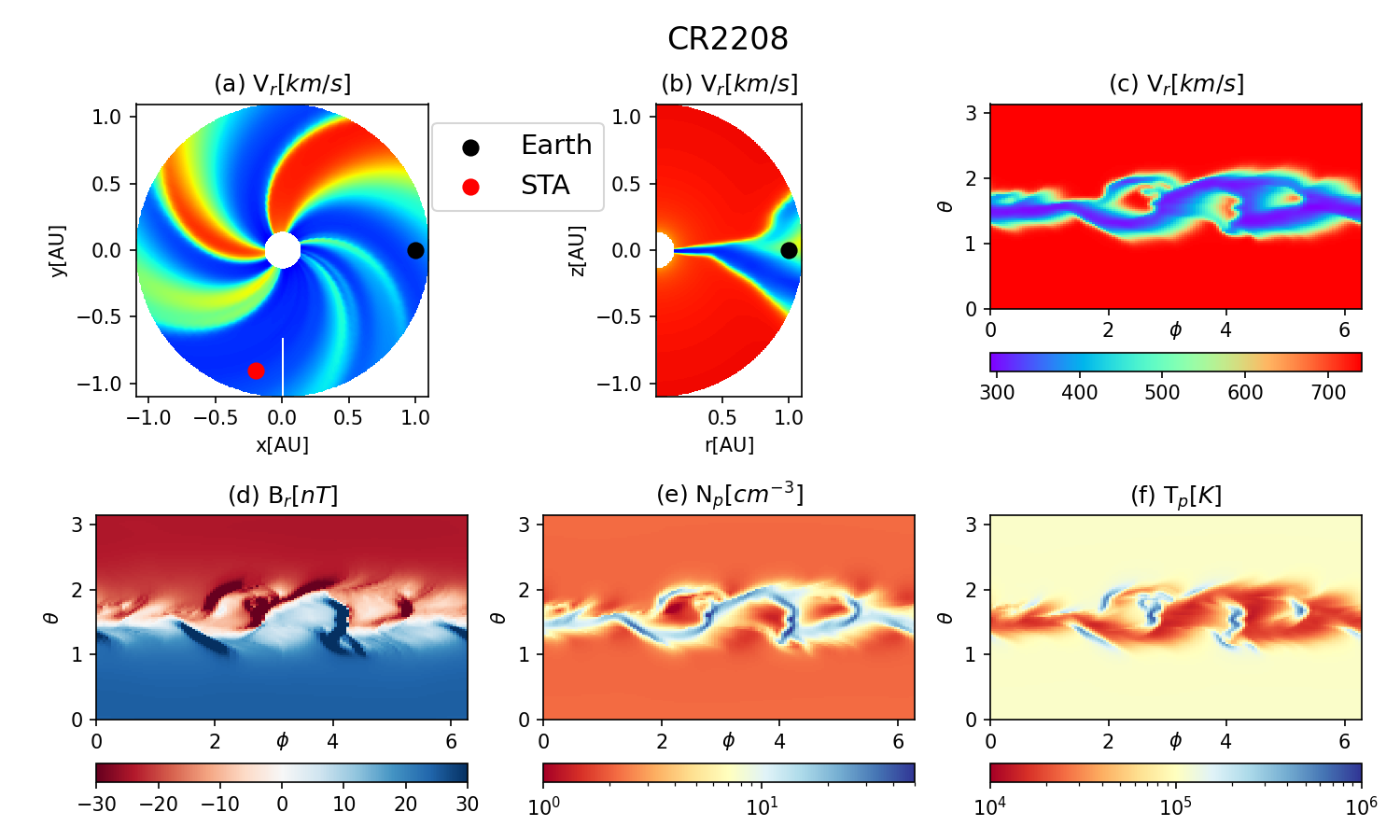}
\centerline{\bf   
      \hspace{0.48\textwidth}  \color{black}{(B)}
         \hfill}
\includegraphics[width=0.65\textwidth]{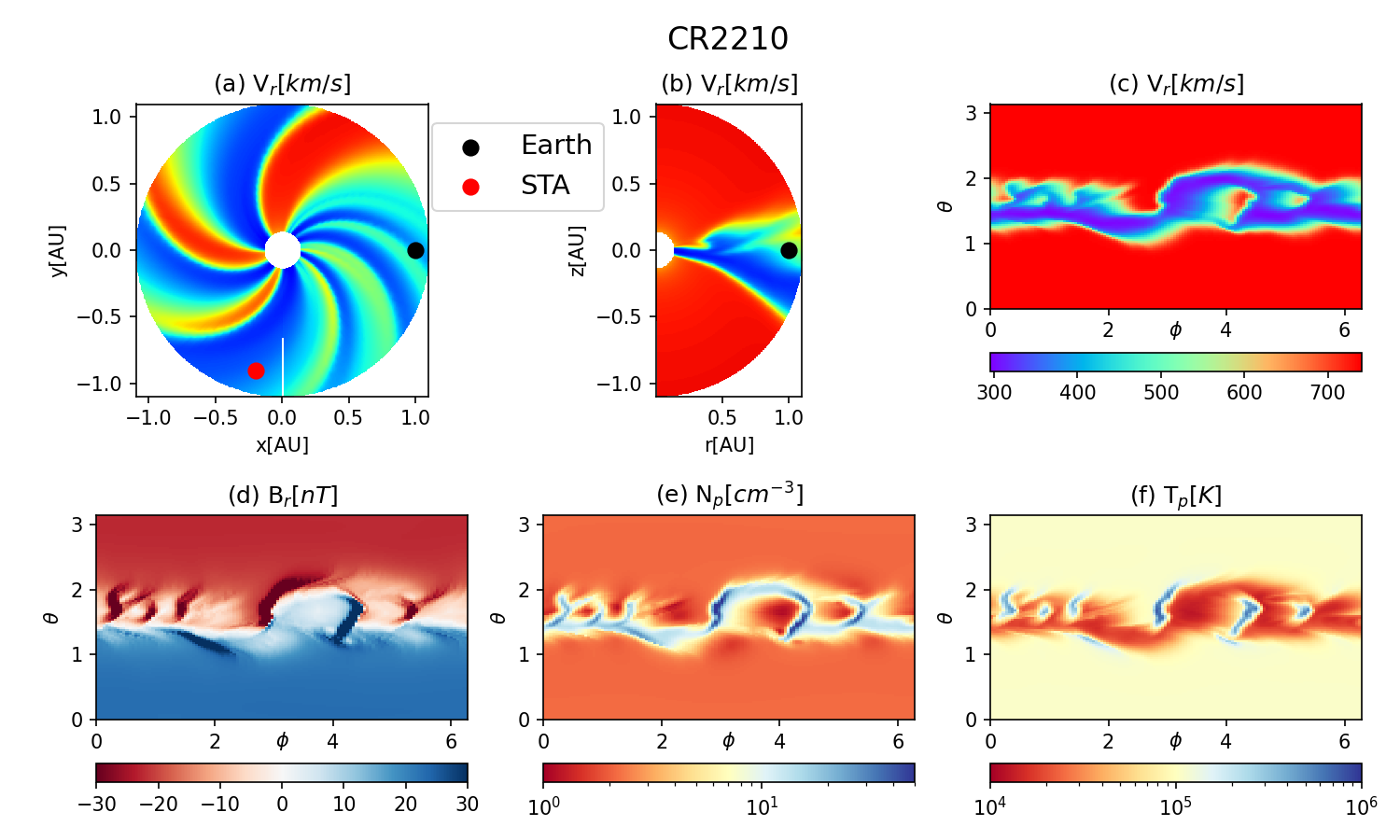}
\centerline{\bf   
      \hspace{0.48\textwidth}  \color{black}{(C)}
         \hfill}
\includegraphics[width=0.65\textwidth]{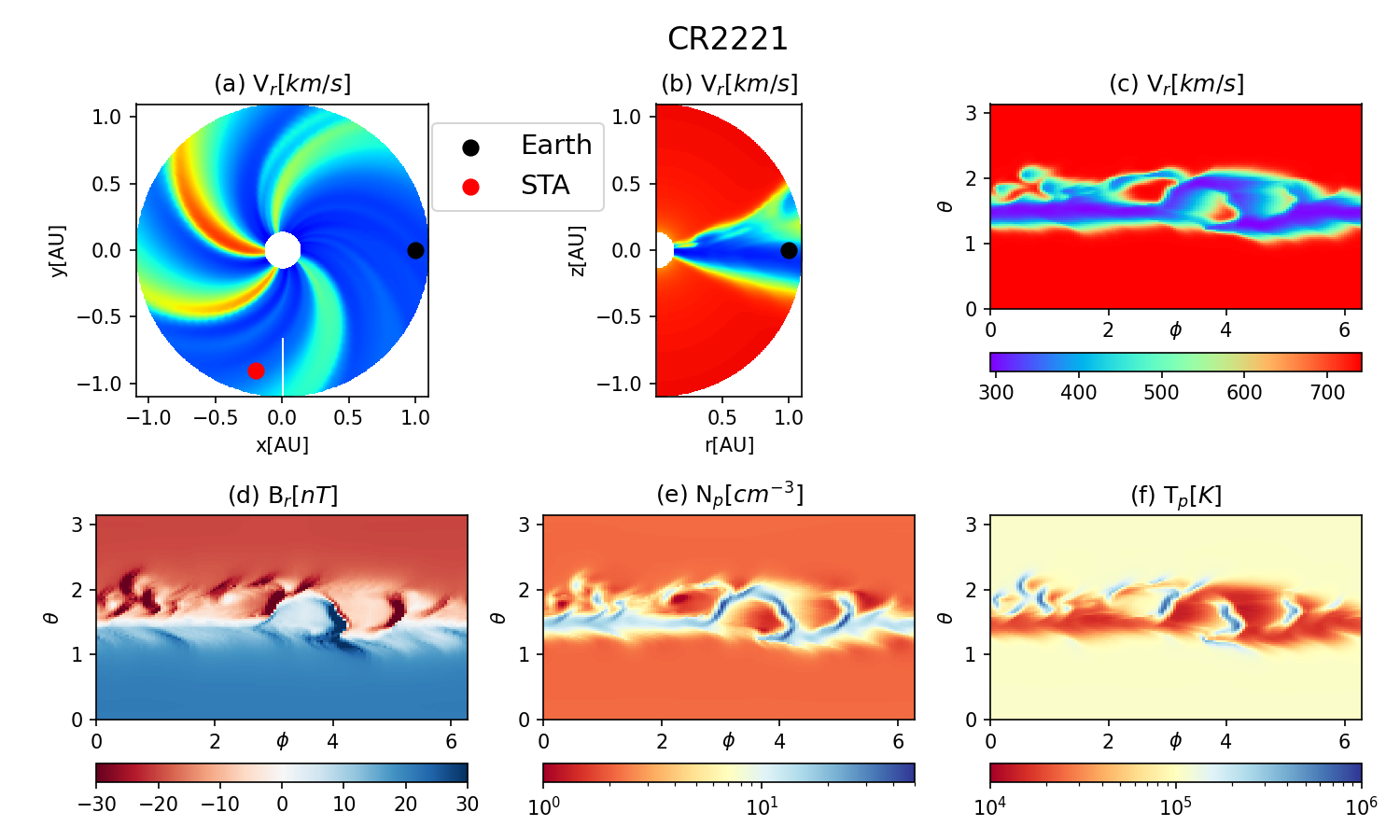}
\caption{Snapshots of the outputs of the inner heliospheric model for CR2208 (A), CR2210 (B), and CR2221 (C). In all the plots, panels (a), (b), and (c) show the radial velocity $V_{r}$, (d) the radial magnetic field $B_{r}$, (e) the proton number density $N_{p}$ and (e) the proton temperature in logarithm scale. Top-left panels are in the $r-\phi$ equatorial plane, top-middle panels are in the $r-\theta$ meridional plane at $0^{\circ}$ longitude and bottom panels are in the $\theta-\phi$ plane at $r=1$ AU. In all panels (a) and (b), the black and red circles represent the projected location of Earth and STA}.
\label{Selected_CRs}
\end{figure}

In Figure \ref{Selected_CRs}, we present snapshots of the output variables for the relaxation, i.e., the steady state of the SW corresponding to a representative set of CRs: CR2208, CR2210, and CR2221. In all panels (a)-(c), we display the radial velocity in km s$^{-1}$ in the equatorial and meridional planes and a slice of constant radial distance at 1 AU, correspondingly. In panels (d)-(f), we show the radial magnetic field in nT, the number density in cm$^{-3}$, and the temperature in Kelvin, respectively. The number density and temperature are shown on a logarithmic scale to discern their structure more clearly. The typical SW structures are visible in the radial velocity cuts, as shown in panels (a)-(c). Mainly, we identify steady-state slow and fast SW streams in the equatorial cuts on panels (a), typically denominated as CIRs. In addition, in all panels (d), the radial magnetic field component shows a polarity change, identified with the heliospheric current sheet (HCS).
Furthermore, in all panels (e), we visualize high-density SW structures near the HCS while low-density SW streams develop in surrounding regions. This behavior is consistent with the radial velocity as schematized in all panels (c), where slow SW streams are located along the HCS. In contrast, fast SW streams dominate at higher latitudes. Also, the fast SW streams have a low density, as expected due to the polytropic thermodynamic relation between the two variables, e.g., \cite{Riley_et_al_2001}. The radial slices of the temperature displayed on panels (f) show regions of lower temperatures near the HCS and higher temperatures in the surrounding regions. Furthermore, in the surrounding regions of the HCS, in the slices at 1 AU, we observe some compression and rarefaction regions that are likely formed due to the interaction of the high-speed streams (HSSs) and the low-speed streams \citep{Gosling_et_al_1972, Gosling&Pizzo_1999}.

\subsection{Comparisons with in-situ measurements of OMNI (Earth) and STA}
\label{Comparisons_with_observations}

We compare the steady-state SW solutions obtained with SR3D-I and SR3D-R with data from OMNI (Earth-based spacecraft) and STA and the MAS results in the inertial frame for all the CRs listed above. To do so, we use the PsiPy tool \citep{David_Stansby_and_Pete_Riley_PsiPy}, which is helpful for reading, processing, and visualizing MHD models developed by Predictive Science Inc. 

\begin{figure}[h]
\centering
\includegraphics[width=0.83\textwidth]{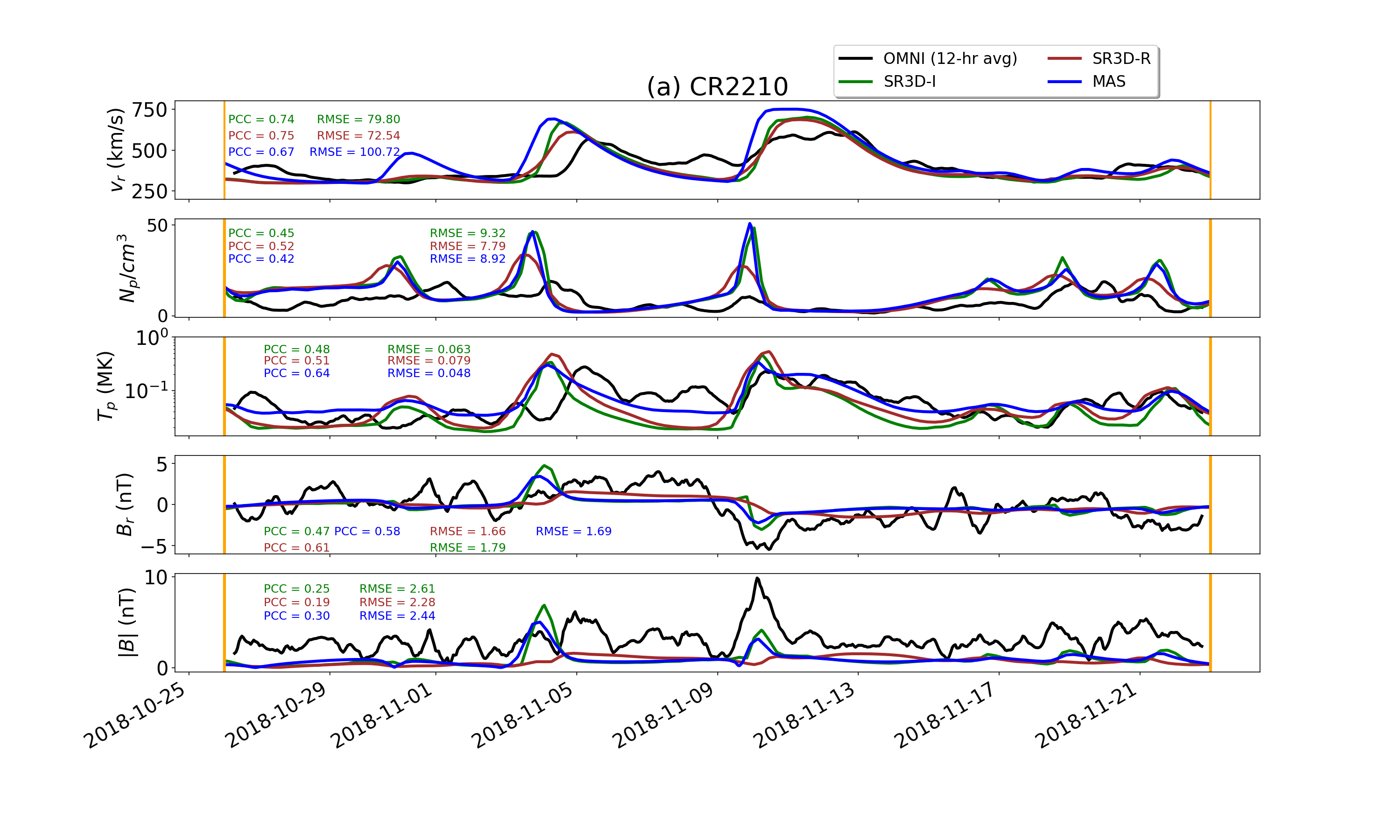}
\includegraphics[width=0.83\textwidth]{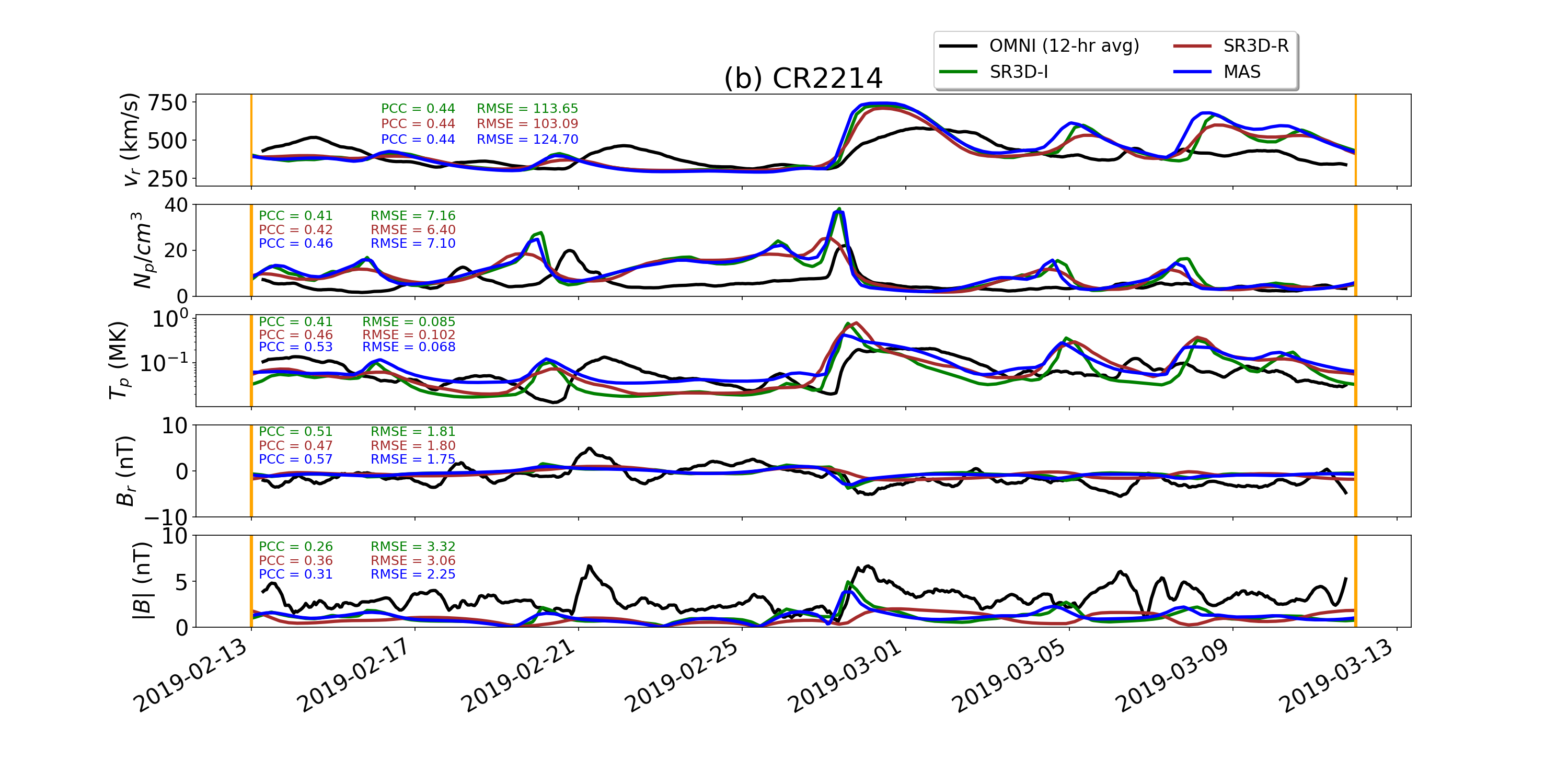}
\includegraphics[width=0.83\textwidth]{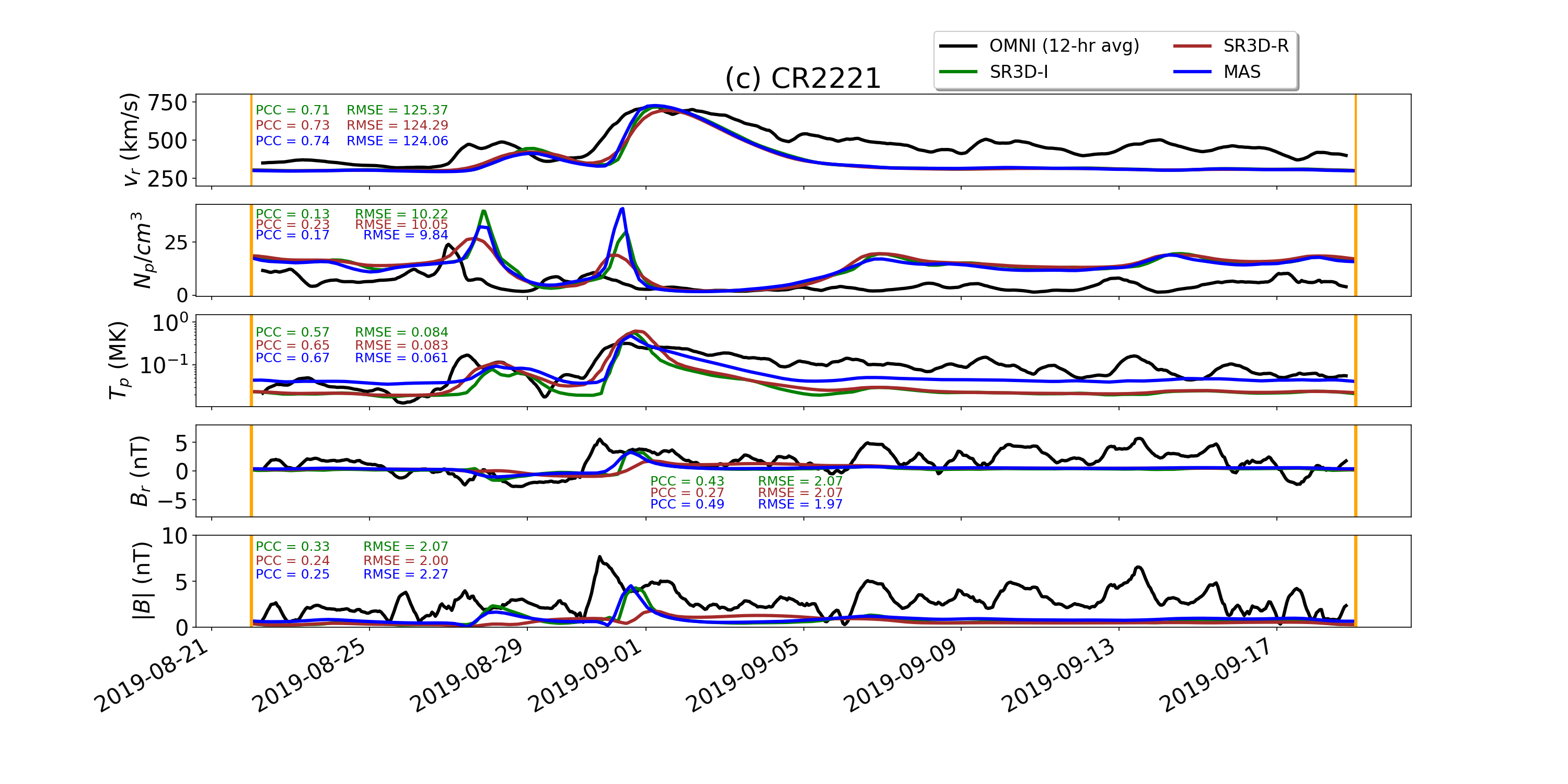}
\caption{Comparison of model results of SR3D-I (green curves), SR3D-R (brown curves), and MAS (blue curves) for the radial velocity $V_{r}$ (km s$^{-1}$), number density $N$ (N/cm$^{3}$), temperature $T$ (MK) in logarithmic scale, radial magnetic field $B_{r}$ (nT) and the magnetic field magnitude $|B|$ (nT) with OMNI in-situ 12-hours averaged measurements (black curves) for CR2210 (a), CR2214 (b), and CR2221 (c).}
\label{Models_vs_OMNI}
\end{figure}

In Figure \ref{Models_vs_OMNI}, we show the results for comparisons between the steady state SW solutions of SR3D-I (green curves) and SR3D-R (brown curves) and MAS (blue curves) with the OMNI in-situ measurements 12-hours averaged (red curves) for three representative CRs: CR2210, CR2214, and CR2221. The 12-hour average means that we make the average 6 hours behind and 6 hours ahead of OMNI in-situ measurements. We include the results of MAS since it has been extensively compared with observations \citep{Riley_et_al_2021}; therefore, MAS serves as a reference model to the SR3D solutions. For example, the comparisons between the models and the radial velocity $V_{r}$ for CR2210 show regions of slow SW ($\sim 300$ km s$^{-1}$) in the first ten days observed by OMNI, which the three models capture. Notably, the three models capture the rise in speed observed at about 2018-11-07; however, there is a slight phase difference. In particular, the three models match the variations from slow to fast wind observed from 2018-11-10 to 2018-11-22. The solutions of SR3D-I and MAS behave similarly, but the SR3D-R results show a slight phase difference compared with the latter models.
Regarding number density, the models overestimated the observed values. It is also evident that SR3D-I and MAS overlap, but SR3D-R solutions are sharper than MAS. SR3D-R captures lower densities at the sharp regions, and the phase shift is visible. For the comparisons with temperature, we see that the three models capture the regions of low temperatures; however, the models overestimate the regions with high temperatures. This behavior is consistent with the results shown in the number density. In the case of the radial magnetic field, we note that the three models only capture global variations, i.e., the changes of sign, but underestimate their amplitude. This behavior has already been reported by, for example, \cite{Riley_et_al_2012, Linker_et_al_2017}, and it can be improved if we multiply the numerical results by a factor of about three. Finally, for CR2210, we display comparisons with the magnetic field magnitude, where it is discernible that the three models underestimate its strength. 

In the middle panels of Figure \ref{Models_vs_OMNI}, we show the results for CR2214. In this case, the slow and fast SW streams are well captured by the three models, as shown in the radial velocity time series. However, the three models overestimate the two rises in velocity occurring at about 2019-02-21 and 2019-03-01, respectively. Again, the slight phase shift between SR3D-I and SR3D-R is visible. In addition, OMNI observations show streams of low density ($\sim 20$ cm$^{-3}$) in most of CR2214; however, the three models overestimated it, and instead, they obtained higher density streams. The comparisons with temperature show the three models capture the global temperature variations. However, they obtain solutions with hotter regions where the observed temperature is lower and colder regions where the in-situ measurements indicate high temperatures. This result is related to the density behavior since the three models overestimated low-density SW streams observed by the in-situ measurements. Generally, low-density regions are hot, so the models struggle to match them, as noticed in the temperature plots. The radial magnetic field and the magnetic field magnitude time series obtained with the three models underestimate the global variations observed by OMNI, similar to the results of CR2210.  

In the bottom panels of Figure \ref{Models_vs_OMNI}, we display the results for CR2221. In this CR, the three models captured the HSSs observed on about 2019-08-30. However, the low density that characterizes these regions was considerably overestimated by SR3D-I and MAS, while SR3D-R solutions describe lower-density HSSs, which are close to the observed values by OMNI. Furthermore, the temperature of the HSSs is colder than the one estimated by the models. Despite this mismatch, the three models can describe the increase in temperature related to the HSSs. Again, the three models underestimate the radial and magnetic fields' magnitude. 

\FloatBarrier

\begin{figure}[h]
\centering
\includegraphics[width=0.83\textwidth]
{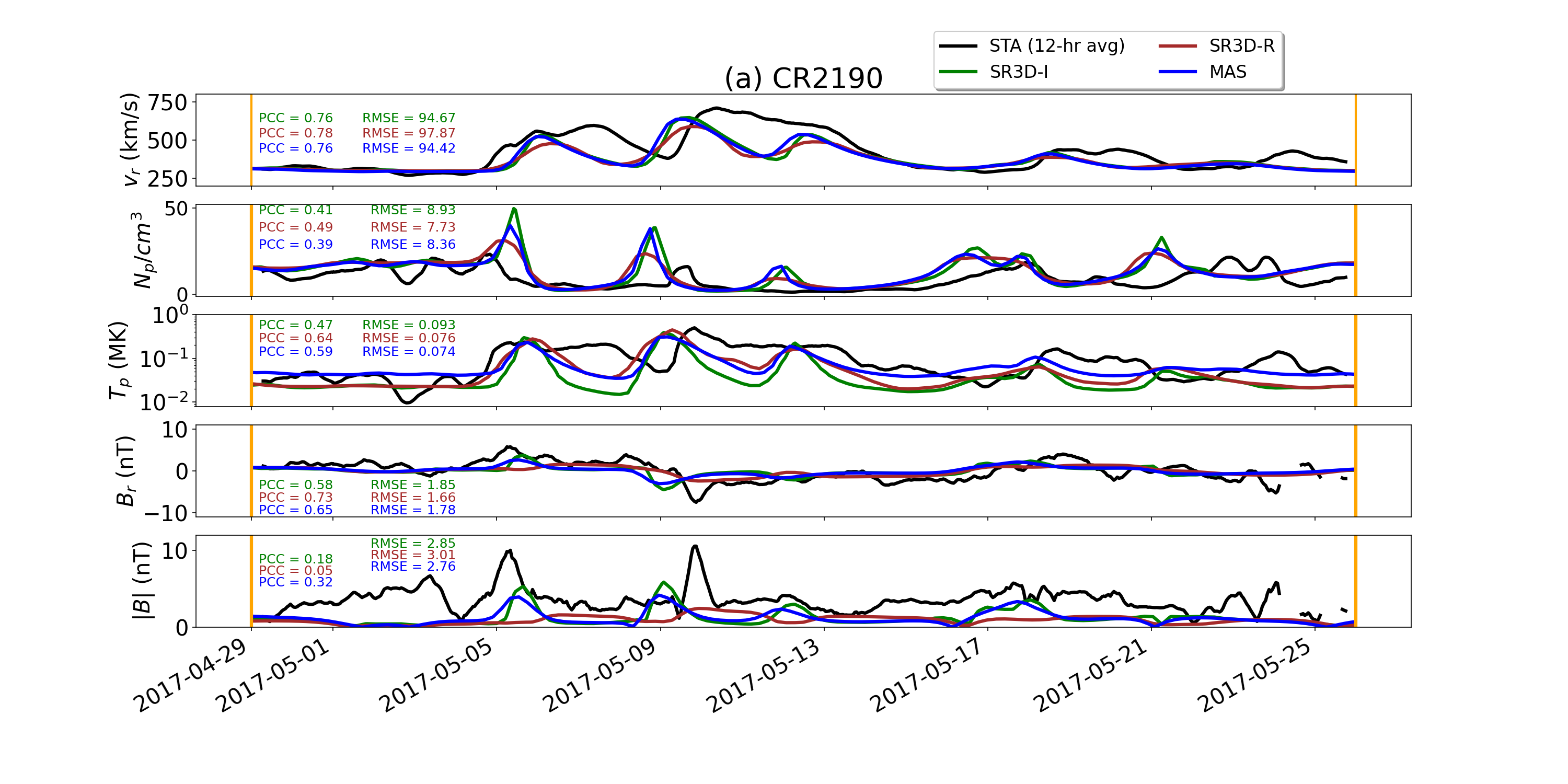}
\includegraphics[width=0.83\textwidth]
{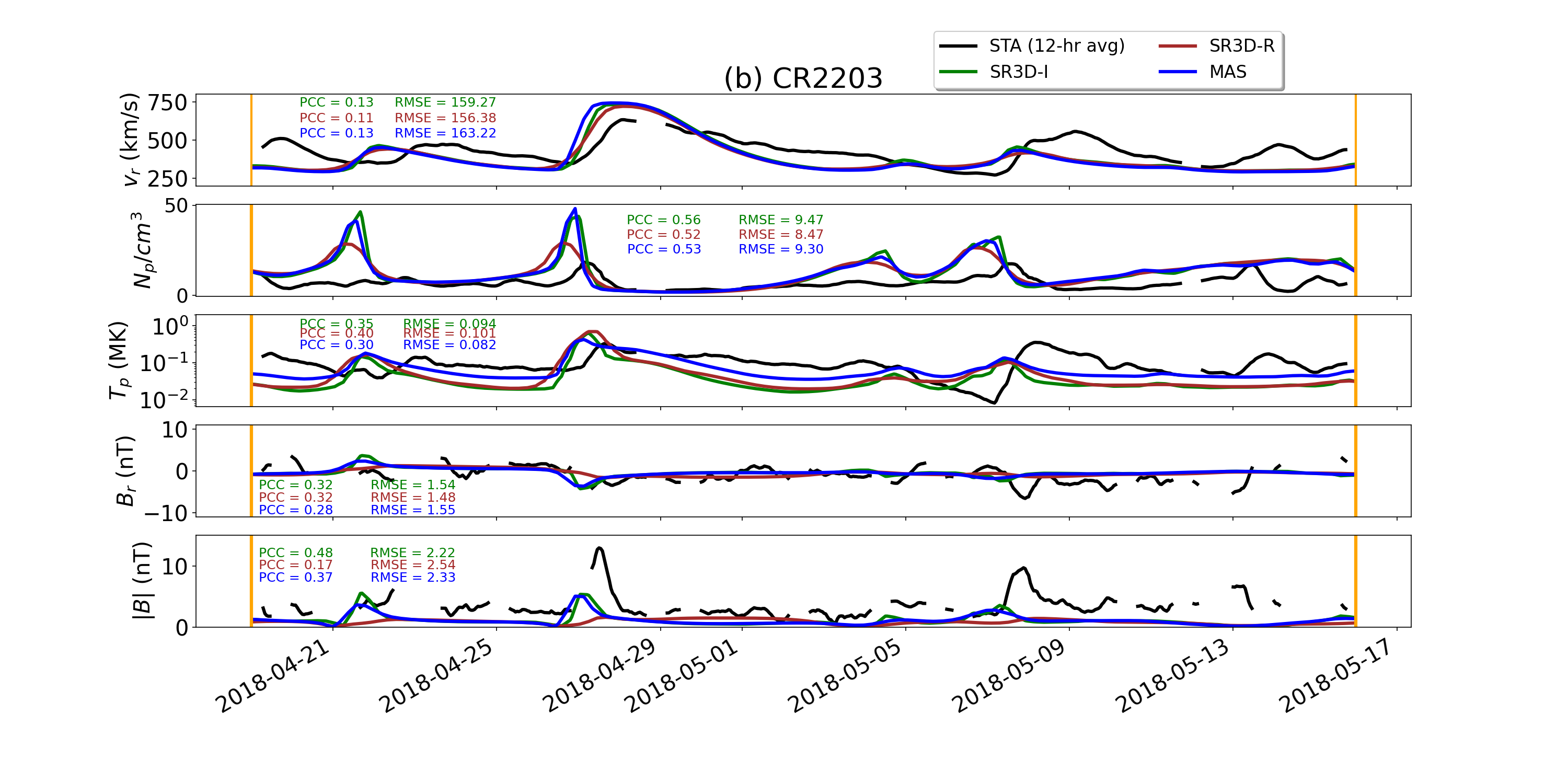}
\includegraphics[width=0.83\textwidth]{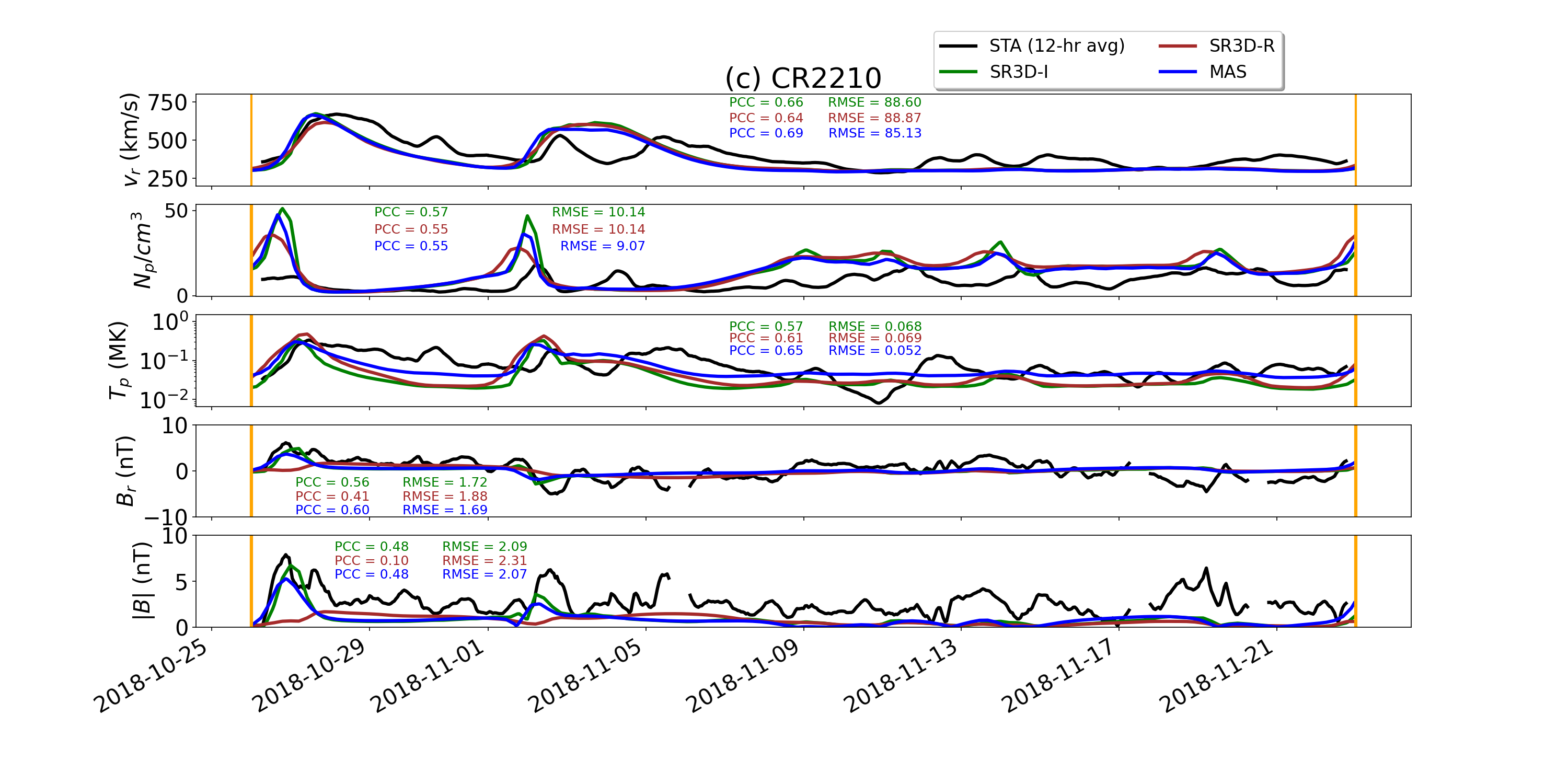}
\caption{Comparison of model results of SR3D-I (green curves), SR3D-R (brown curves) and MAS (blue curves) for the radial velocity $V_{r}$ (km s$^{-1}$), number density $N$ (N/cm$^{3}$), temperature $T$ (MK) in logarithmic scale, radial magnetic field $B_{r}$ (nT) and the magnetic field magnitude $|B|$ (nT) with STA in-situ 12-hours averaged measurements (black curves) for CR2190 (a), CR2203 (b) and CR2210 (c).}
\label{Models_vs_STEREO-A}
\end{figure}

In Figure \ref{Models_vs_STEREO-A}, we show the results for comparisons between the model solutions and STA in-situ measurements 12 hours averaged for the case of CR2190, CR2203, and CR2210. In the case of CR2190 (top panels), the three models matched the slow SW streams but underestimated the high-speed SW streams. For the number density, we note that the three models overestimated the low-density streams; however, they reproduce the rise. Again, we identify that three models underestimated the magnitude of the magnetic field. In the middle panels, we show the comparisons for the CR2203. Here, the three models match the radial velocity of the SW streams. Again, they overestimated the low-density SW streams for the number density, but they captured the large-scale increases and decreases. In the case of the radial magnetic field and the magnetic field magnitude, the three models underestimated the strength; however, they describe the global behavior of the magnetic field. In the case of CR2210 (bottom panels), STA observed three CIRs as reported by \cite{Allen_et_al_2021}. The first CR was observed between 2018-10-26 at 06:30 and 2018-10-27 at 20:50; the second started at about 06:00 of 2018-11-02 and finalized at 18:10 of 2018-11-02; while the third one was observed between 2018-11-22 at 20:40 and 2018-11-23 at 00:35. From the three CIRs, the models match reasonably well the radial velocity and temperature for the first CIR. The match to the second CIR is acceptable; however, the three models did not match the third CIR. Notably, the three models underestimated the radial velocity and overestimated the density of the CIR.

\FloatBarrier

In all the time series of Figures \ref{Models_vs_OMNI} and \ref{Models_vs_STEREO-A}, we identify slight differences regarding the results obtained with SR3D-R, SR3D-I, and MAS. In particular, we note a difference in the sharpness of features and a phase shift between both solutions. These differences could be related to how we solve the MHD equations in each frame of reference. However, the differences could be related to other reasons, as discussed in more detail in Appendix \ref{Appendix_A}. 

Furthermore, from the three model comparisons with OMNI and STA in-situ measurements as described in the above paragraphs and Figs. \ref{Models_vs_OMNI} and \ref{Models_vs_STEREO-A}, we note that MAS results differ from SR3D-I and SR3D-R results. It is appropriate to compare MAS and SR3D-I solutions since both models solve the same MHD equations in the inertial frame and employ identical boundary conditions. At the same time, SR3D-R solves the MHD equations in the rotating frame, accounting for centrifugal and Coriolis forces. This introduces additional numerical diffusion, as elaborated further in Appendix \ref{Appendix_A}. Focusing on the comparison between SR3D-I and MAS, both models employ different numerical algorithms; for instance, MAS uses the finite differences method to discretize the equations, and SR3D-I employs the finite volume discretization and high-resolution-capturing methods to solve the MHD equations, which are the algorithms used by PLUTO. Besides, MAS evolves the magnetic field through a magnetic vector potential, which maintains the $\nabla\cdot{\bf B}$ condition to a round-off value.
In contrast, SR3D-I uses Powell's eight-wave formulation that maintains $\nabla\cdot{\bf B}$ to a truncation order error. Based on the differences above, the main reason for the disparities between MAS and SR3D-I solutions is mainly related to the numerical algorithms used to solve the MHD equations. Commonly, finite difference methods produce more diffusion than finite volume solutions for MHD equations. So, based on the explanation above, the differences between MAS and SR3D-I solutions relate to numerical artifacts.

\subsubsection{Statistical analysis}
\label{Statistical_analysis}

We performed a statistical analysis to validate the quality of the three models, SR3D-I, SR3D-R, and MAS, compared to in-situ measurements. In particular, we compared measurements and model solutions by estimating the Root Mean Square Error (RMSE) and the Pearson Correlation Coefficient (PCC). The RMSE represents the mean squared difference between measurements and models and is less sensitive to atypical values \citep[see, e.g.,][]{REISS2022}. At the same time, the PCC measures the linear correlation between two data sets. These two statistical measures are defined as follows:

\begin{eqnarray}
\textrm{RMSE} &=& \sqrt{\frac{1}{N}\sum_{k=1}^{N}(f_{k}- O_{k})^{2}}, \label{RMSE} \\
\textrm{PCC} &=& \frac{\sum_{k=1}^{N}(f_{k}-\bar{f_{k}})^{2}(O_{k}-\bar{O_{k}})^{2}}{\sum_{k=1}^{N}(f_{k}-\bar{f_{k}})^{2}\sum_{k=1}^{N}(O_{k}-\bar{O_{k}})^{2}}, \label{PCC} 
\end{eqnarray}

\noindent where $f_{k}$ and $O_{k}$ are the $k-th$ elements of $N$ total model values and in-situ observations, respectively. The $\bar{f_{k}}$ and $\bar{O_{k}}$ in equation \ref{PCC} represent the mean values. Despite the small shifts in time between simulation and data that could lead to misleadingly poor performance using to point-to-point comparisons, more sophisticated techniques, such as dynamic time warping \citep[see, e.g.,][]{10.1093/mnras/stab2512, Samara_et_al_2022}, are beyond the scope of this paper.

\begin{figure*}
\centering
\includegraphics[width=0.43\textwidth]{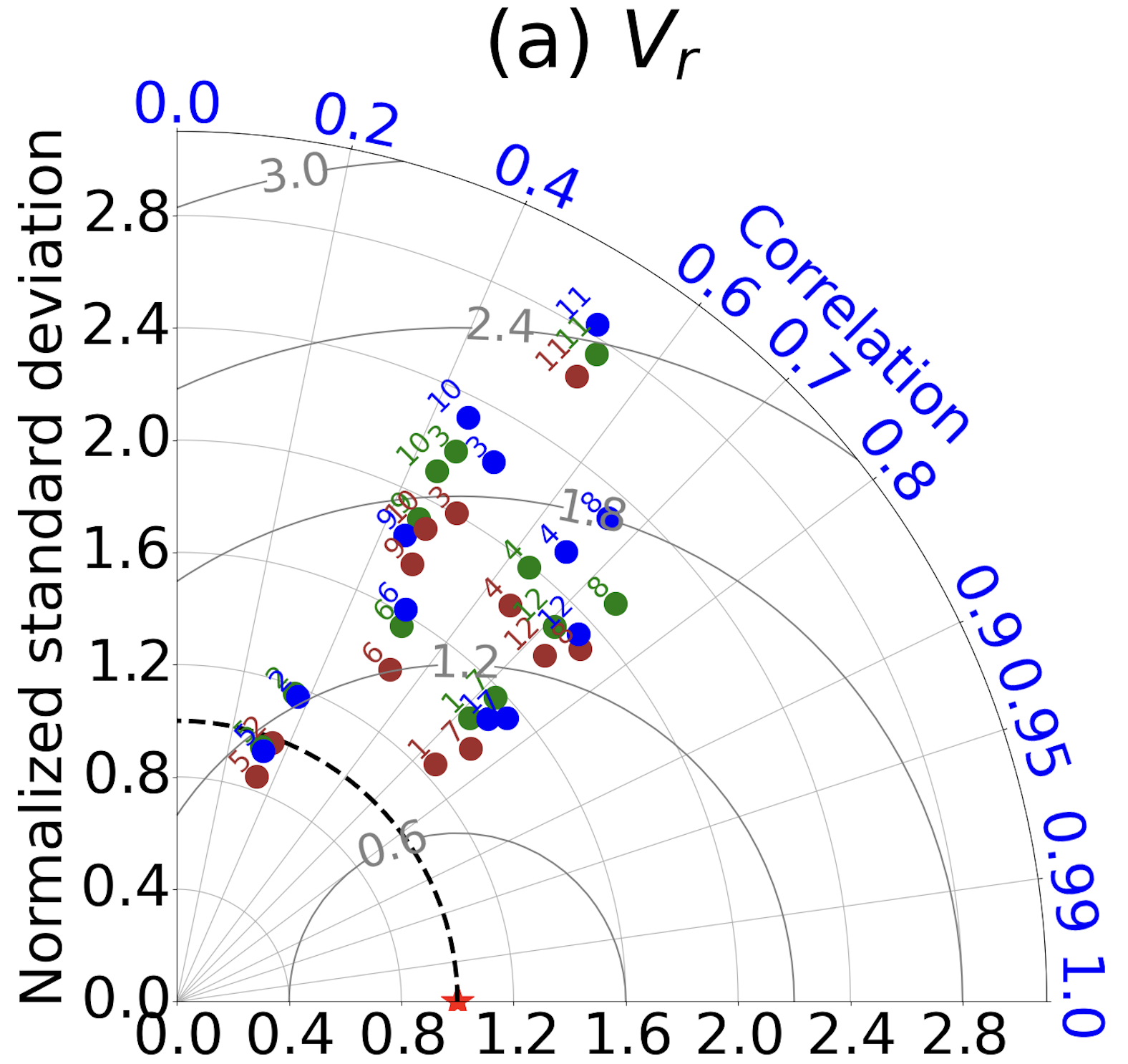}
\includegraphics[width=0.50\textwidth]{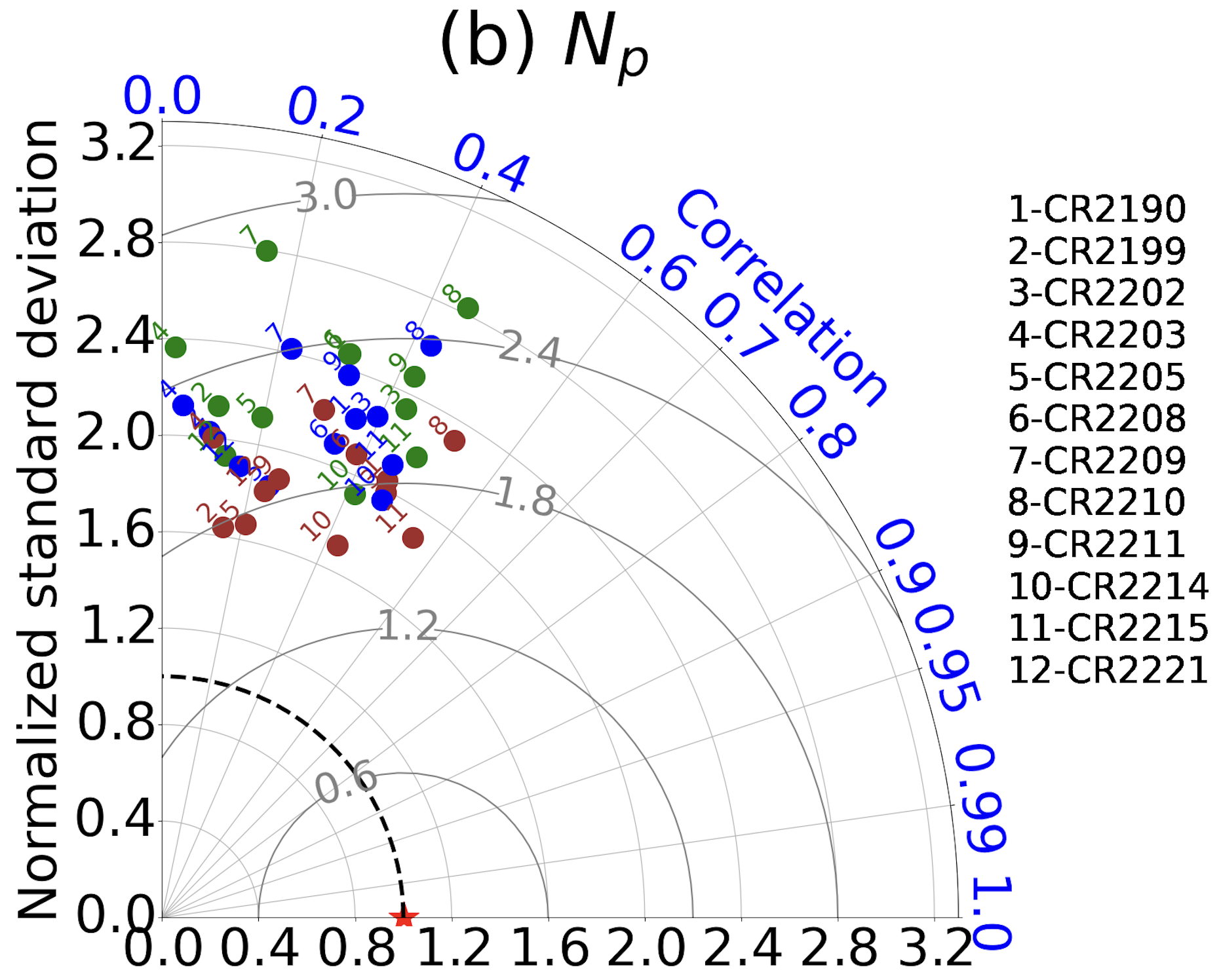}
\includegraphics[width=0.45\textwidth]{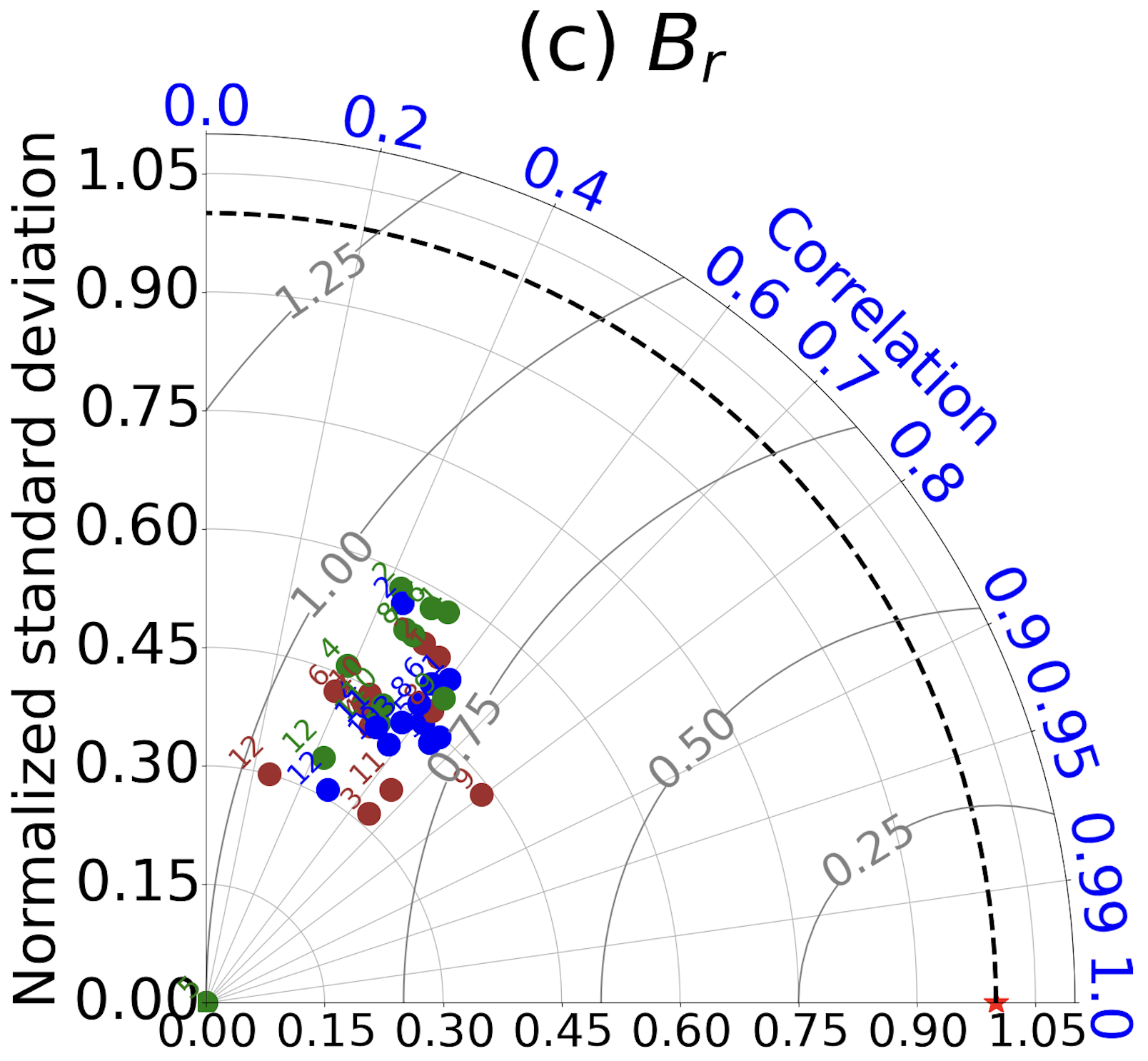}
\includegraphics[width=0.5\textwidth]{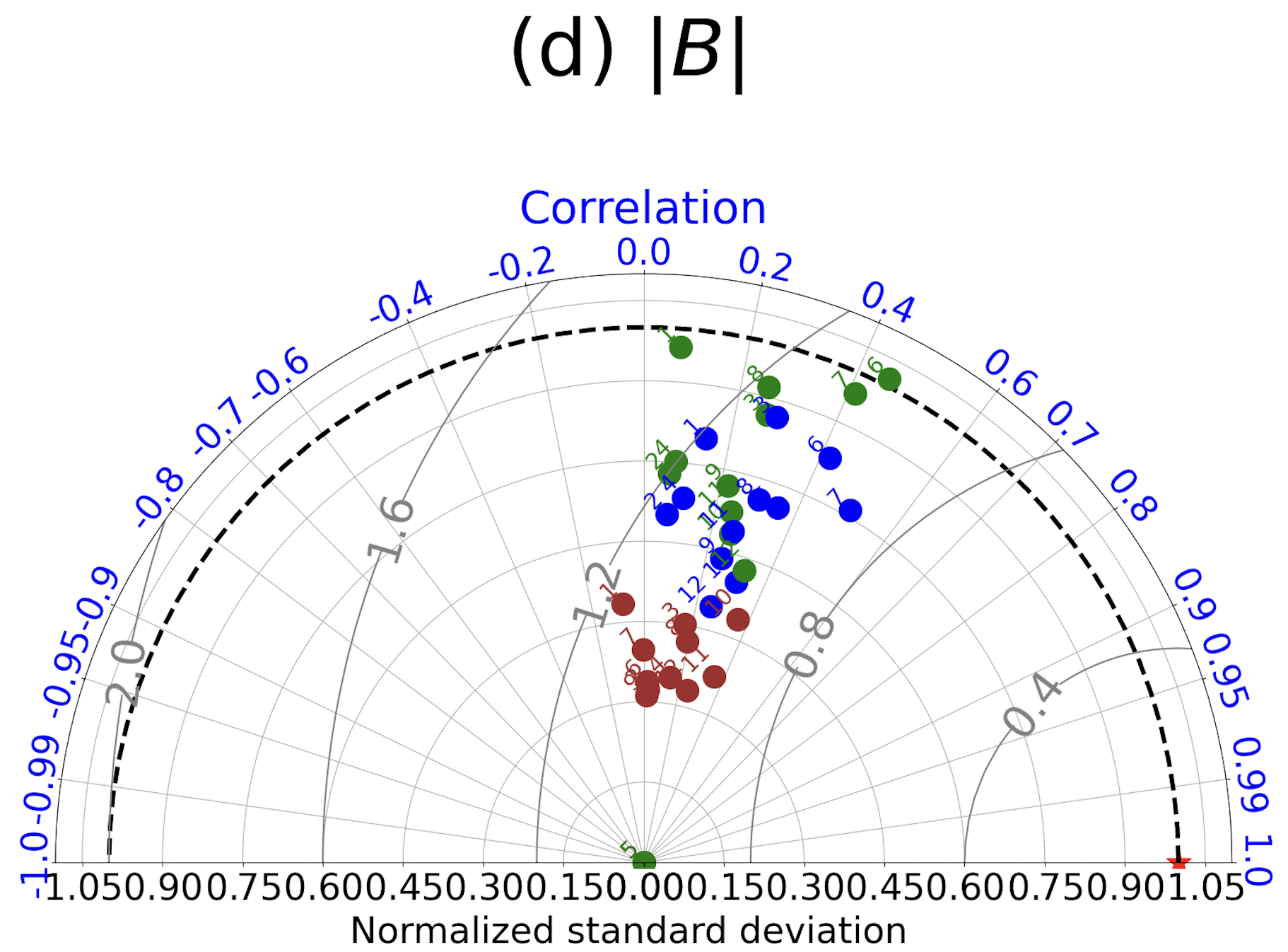}
\includegraphics[width=0.15\textwidth]{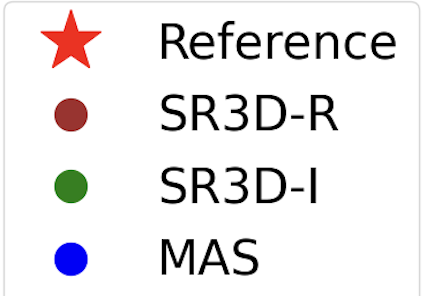}
\caption{Taylor diagrams to models with OMNI 12-hr avg measurements for all CRs corresponding to $V_{r}$ (a), $N_{p}$ (b), $B_{r}$ (c), and $|B|$ (d).}
\label{Taylor_diagrams_OMNI}
\end{figure*}

In Tables \ref{tab:Stat_models_vs_OMNI} and \ref{tab:Stat_models_vs_STEREO-A} of Appendix \ref{Appendix_B}, we show all the statistical results in terms of RMSE and PCC for OMNI and STA, correspondingly. To avoid over-information, we use Taylor diagrams \citep{Taylor_2001}, combining correlation, centered RMSE difference, and variances to summarize all the statistics and discern the best results for all the SW variables and all CRs mentioned above. Note that there is a difference between RMSE shown in Tables \ref{tab:Stat_models_vs_OMNI} and \ref{tab:Stat_models_vs_STEREO-A} with the centered RMSE difference shown in the Taylor diagrams. In particular, the only difference is that before computing the RMSE, the average values of the data and model are first subtracted from the former. These diagrams are helpful to validate multiple aspects in a single diagram, and they have already been applied to the validation of ambient solar wind numerical models against in-situ measurements \citep[see, e.g.,][]{Riley_et_al_2013, Owens_2018, Reiss_et_al_2020,10.3389/fspas.2020.572084}. The main objective of Taylor diagrams is to recognize the geometric relationship between the correlation coefficient, the RMSE, and the amplitude of variations in the modeled and reference time series. In all the panels of Figure \ref{Taylor_diagrams_OMNI}, the azimuthal position indicates the PCC, the radial distance from the circle at $x$-axis is proportional to the centered RMSE difference, and the distance from the origin is proportional to the normalized standard deviation. Therefore, models with good agreement with in-situ measurements should be close to the red star on the $x$-axis, indicating a similar standard deviation, a high correlation, and a low centered RMSE difference. In particular, we normalize the standard deviation in terms of the lowest deviation estimated for observations in each CR; therefore, the standard deviation values of the models can be less or greater than one in all the cases. Also, we set the observation standard value for all cases equal to one, representing the reference value. This normalization helps tighten all the values with similar scales and compares all the model results more quickly. Furthermore, we set points with three different colors corresponding to each model (SR3D-R, SR3D-I, and MAS) and labeled each point with a number from 1 to 12; these labels correspond to the 12 CRs analyzed. This way, it is easier to identify which of the three models for each CRs is closest to the observation values.

For example, in panel (a) of Figure \ref{Taylor_diagrams_OMNI}, we show the Taylor diagram for the radial velocity $V_{r}$ corresponding to the comparisons with OMNI 12-hr averaged data. In this panel, it is discernible that the best performance of models is for CR2199 and CR2205, whose values are close to the observation data, while CR2215 achieves the worst match of the three models. Globally, from the diagram, MAS results are better than SR3D-I and SR3D-R results for matching the radial velocity. In panel (b), the distribution of the points for proton density $N_{p}$ shows that three models barely match with in-situ measurements for most of the CRs; however, MAS matches acceptable to the observation values for CR2199 and CR2205. In summary, the diagram for proton density shows a clustering of the models for most of the CRs; therefore, they are comparable. The distribution of the model results for the radial magnetic field $B_{r}$, as displayed in panel (c), shows that the three models are close to each other and are close to observation values; MAS shows an appropriate performance for CR2211. Finally, in panel (d), for the magnetic field magnitude $|B|$, we identify that the results are generally poor compared to the other variables. Specifically, in this panel, we see SR3D-R has a negative PCC for CR2190 and small values for CR2199, CR2208, and CR2211. Despite the latter results in panel (d), the three models reach comparable results; mainly, SR3D-I and MAS perform better than SR3D-R.

\begin{figure}
\centering
\includegraphics[width=0.5\textwidth]{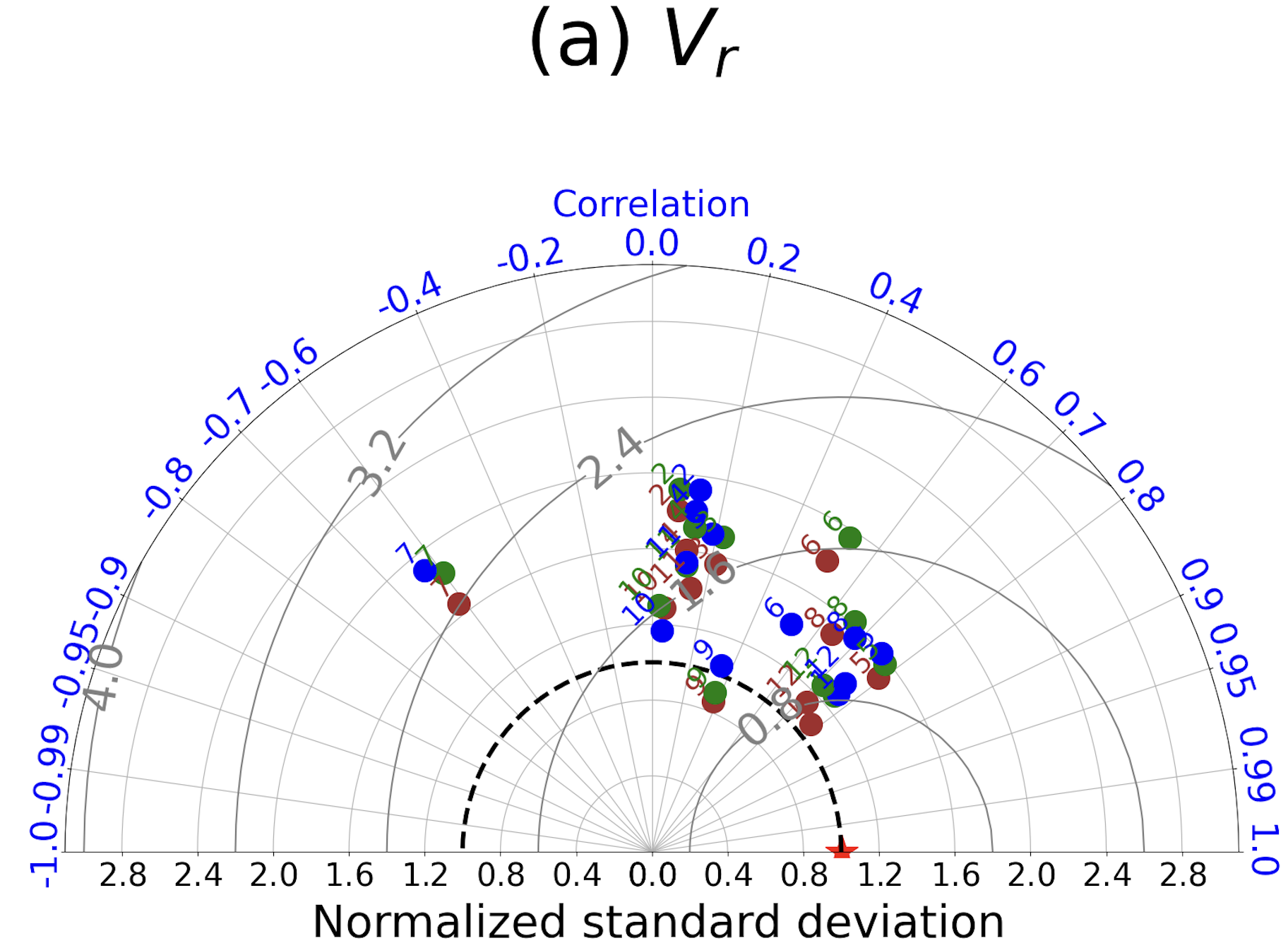}
\includegraphics[width=0.45\textwidth]{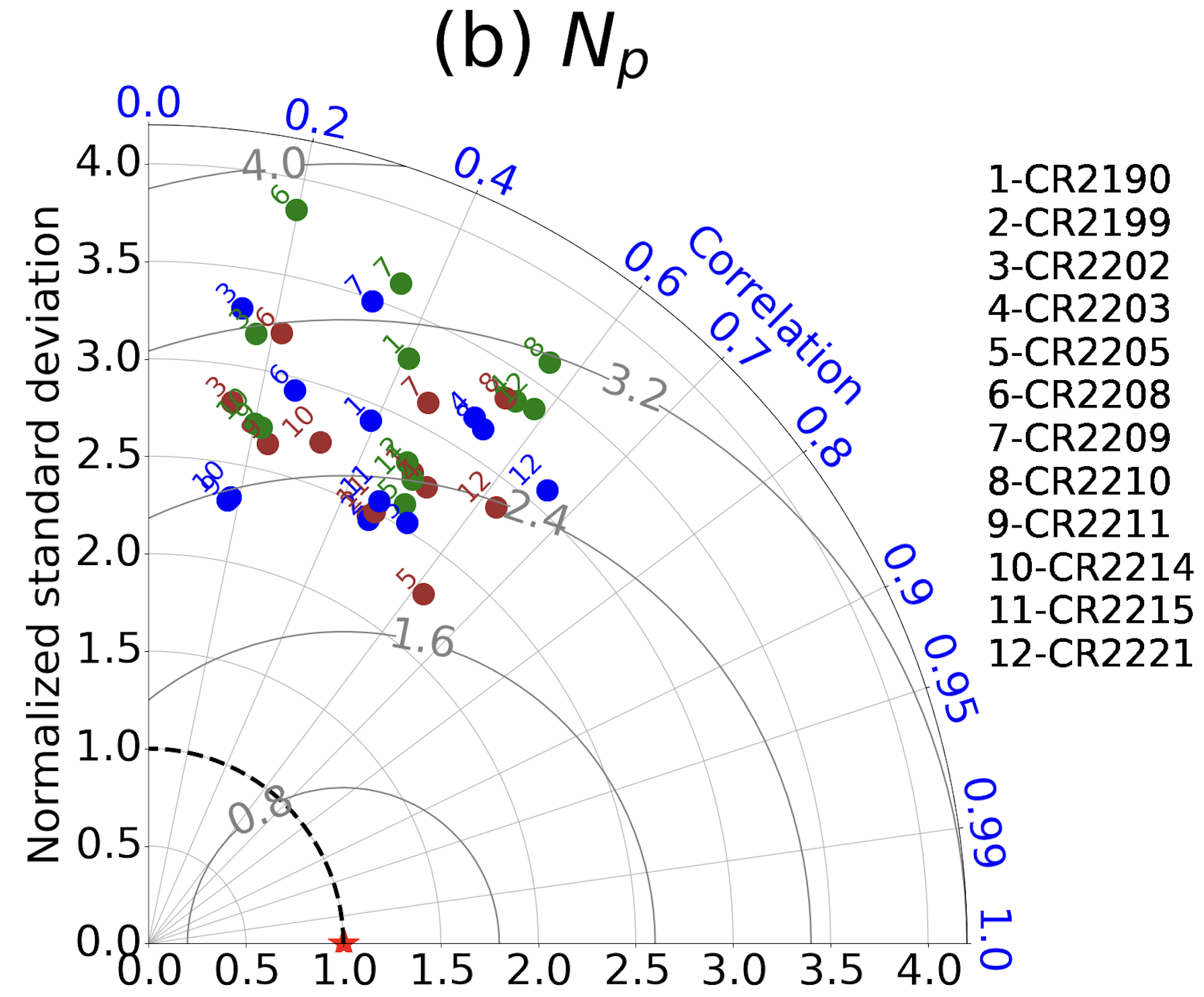}
\includegraphics[width=0.45\textwidth]{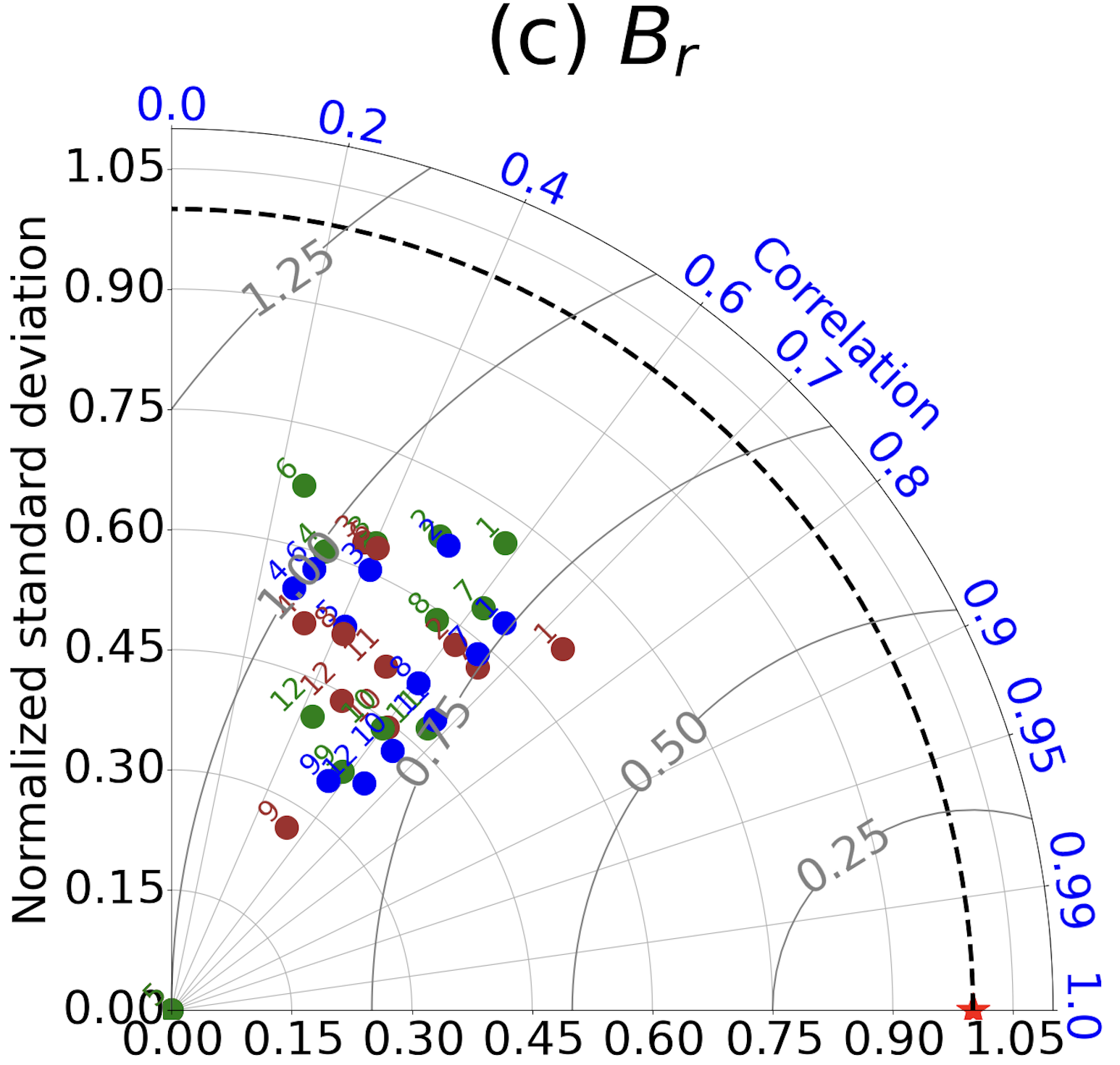}
\includegraphics[width=0.5\textwidth]{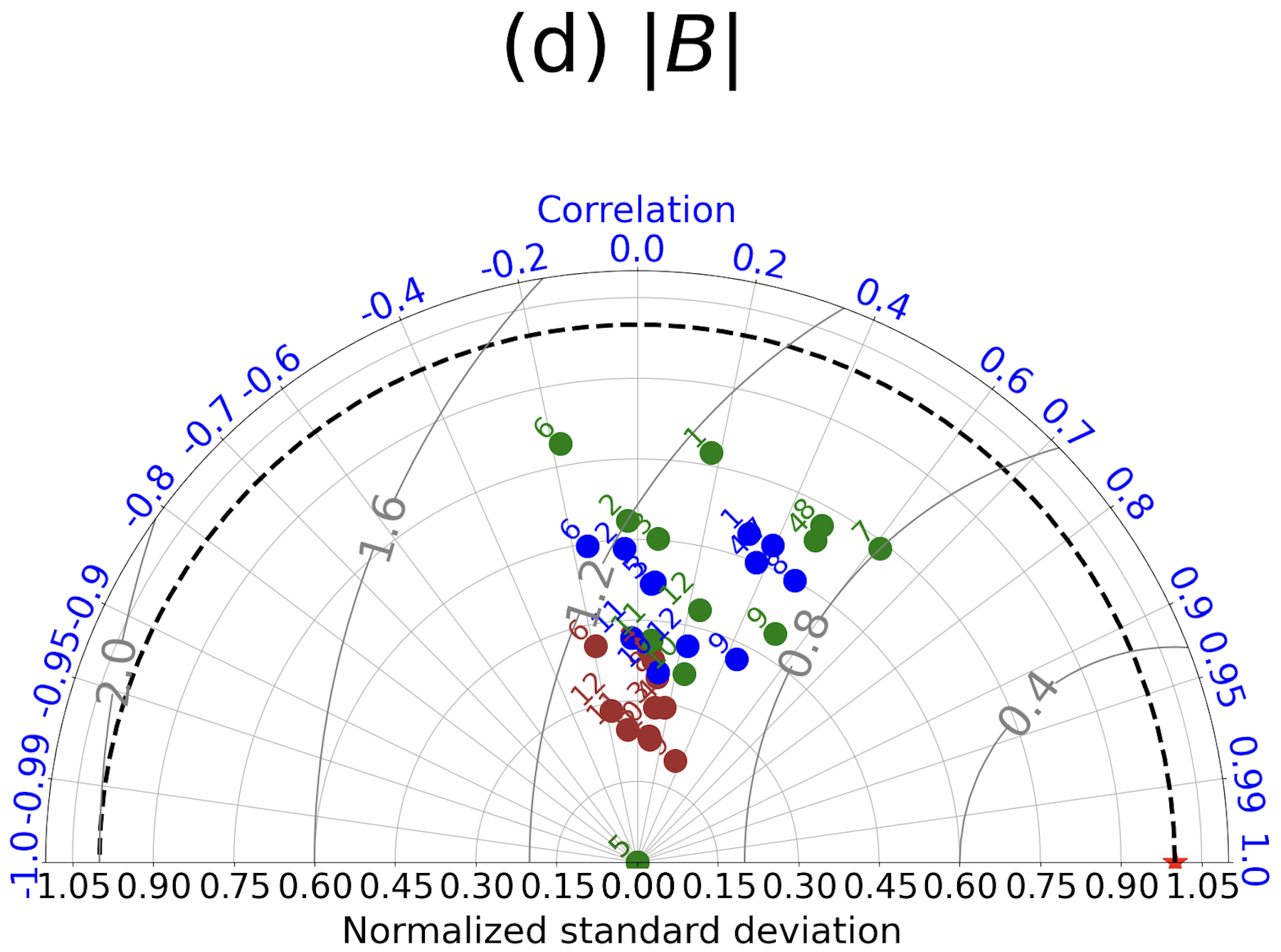}
\includegraphics[width=0.15\textwidth]{Label_OMNI_Taylor.png}
\caption{Taylor diagrams to compare models with STA 12-hr avg measurements for all CRs corresponding to $V_{r}$ (a), $N_{p}$ (b), $B_{r}$ (c), and $|B|$ (d).}
\label{Taylor_diagrams_STA}
\end{figure}

Similarly to the analysis performed for OMNI, we also show Taylor diagrams to summarize the statistical comparisons between models and STA-12 hrs averaged in-situ measurements. In panel (a) of Figure \ref{Taylor_diagrams_STA}, we display the results for the radial velocity $V_{r}$, where the three models show an acceptable performance for most of the CRs, except for CR2209, where three models have negative PCC and high-centered RMSE difference values. At the same time, SR3D-R is closest to the observed radial velocity for CR2190. The proton number density $N_{p}$ displayed in panel (b) indicates that the three models are not close to observations. However, SR3D-R for CR2205 matches reasonably well with STA in-situ measurements. For the magnetic radial field, shown in panel (c), the three models cluster similarly, and they lie between 0.75 and 1.00 centered RMSE difference and between 0.6 and 0.7 PCC, which are consistent with the results obtained in \cite{Riley_et_al_2015}. Lastly, in panel (d), the Taylor diagram for the magnetic field strength shows that SR3D-R has negative PCC values for CR2199, CR2205, CR2208, CR2215, and CR2221. However, the three models reach values reasonably close to the reference magnetic field strength observed by STA.

\FloatBarrier

\FloatBarrier

\subsection{Comparison with similar existing model}
\label{SWASTi_section}

In this work, we have also conducted a comparative study with another existing model that utilizes the PLUTO code, mainly focusing on the impact of initial boundary conditions on the final results at 1AU. This feature involved the application of various coronal models including GONG (\url{https://gong.nso.edu/data/magmap/}) and Air Force Data Assimilative Photospheric Flux Transport (ADAPT: \url{https://nso.edu/data/nisp-data/adapt-maps/}) based Wang-Sheeley-Arge (WSA), as well as HMI-based MAS, integrated with the SWASTi solar wind module \citep{Mayank_2022}. The SWASTi framework, recently developed to forecast and assess the solar wind and CME characteristics within the inner heliosphere, operates on the ideal MHD module of the PLUTO code. The framework also incorporates gravity and adopts an ideal equation of state with $\gamma$ = 1.5. Notably, SWASTi offers a high degree of flexibility regarding time cadence, allowing for intervals as brief as 5 minutes or less, depending on user specifications. This flexibility is facilitated through a virtual spacecraft approach, which stores plasma properties in real-time simulations. Such a generic approach is crucial to understanding the dynamic behavior of heliospheric transients, particularly CMEs \citep[see, e.g.,][]{mayank2023}.

In this study, we have employed the SWASTi's default numerical setup, wherein the form of the WSA speed relation is the follows:

\begin{equation}\label{eq:WSA}
   V_{WSA} = v_{1} + \frac{v_{2}}{(1+f_s)^{\frac{2}{9}}} \times \Bigg(1.0 - 0.8\,exp \Bigg(-\bigg(\frac{d}{w}\bigg)^{\beta}\Bigg)\Bigg)^{3}
\end{equation}

\noindent where $v_1$, $v_2$, and $\beta$ are independent parameters whose values are taken to be 250 km/s, 675 km/s and 1.25, respectively \citep{Narechania_et_al_2021}. $f_s$ and $d$ are the flux tube's areal expansion factor and the foot-point's minimum angular separation from the coronal hole boundary. In contrast to the original WSA relation, the parameter $w$ is not independent; it is determined by the median of the $d$ values corresponding to the field lines that reach Earth's location. This means the adapted speed relation operates with fewer independent parameters, offering a more adaptable approach than the initial WSA relation. Additionally, the MHD domain range spans from 0.1 AU to 2.1 AU radially, -60\textdegree $\,$ to 60\textdegree $\,$ latitudinally, and 0\textdegree $\,$ to 360\textdegree $\,$ longitudinally, with respective grid resolutions of $150\times120\times360$. For further details, readers may refer to \cite{Mayank_2022}.

To clearly understand the differences between SR3D and SWASTi, we have encapsulated the unique features of both models in Table 1. We specify the supported magnetograms and the models employed in the coronal region. Then, it is discernible that both models use different semi-empirical coronal models and magnetograms but employ PLUTO as the heliospheric model. Hence, the purpose of having both models (SWASTi and SR3D) in this subsection is to compare each other and explore their capabilities of capturing solar wind properties, for example, at about 1 AU.

\begin{table}
\begin{center}
\caption{Features of SWASTi and SR3D.}
\label{tab:SWASTi_SR3D}
\begin{tabular}{|c|c|c|c|} 
\hline
Model & Coronal Region & Magnetogram & Heliosphere \\
\hline
SWASTi \citep{Mayank_2022} & WSA & Any (GONG, ADAPT, HMI) & PLUTO \\ 
SR3D (this work) & MAS & HMI & PLUTO \\ 
\hline
\end{tabular}
\end{center}
\end{table}

\begin{figure}
\centering
\centerline{\bf   
      \hspace{0.49\textwidth}  \color{black}{(a)}
         \hfill}
\includegraphics[width=0.8\textwidth]{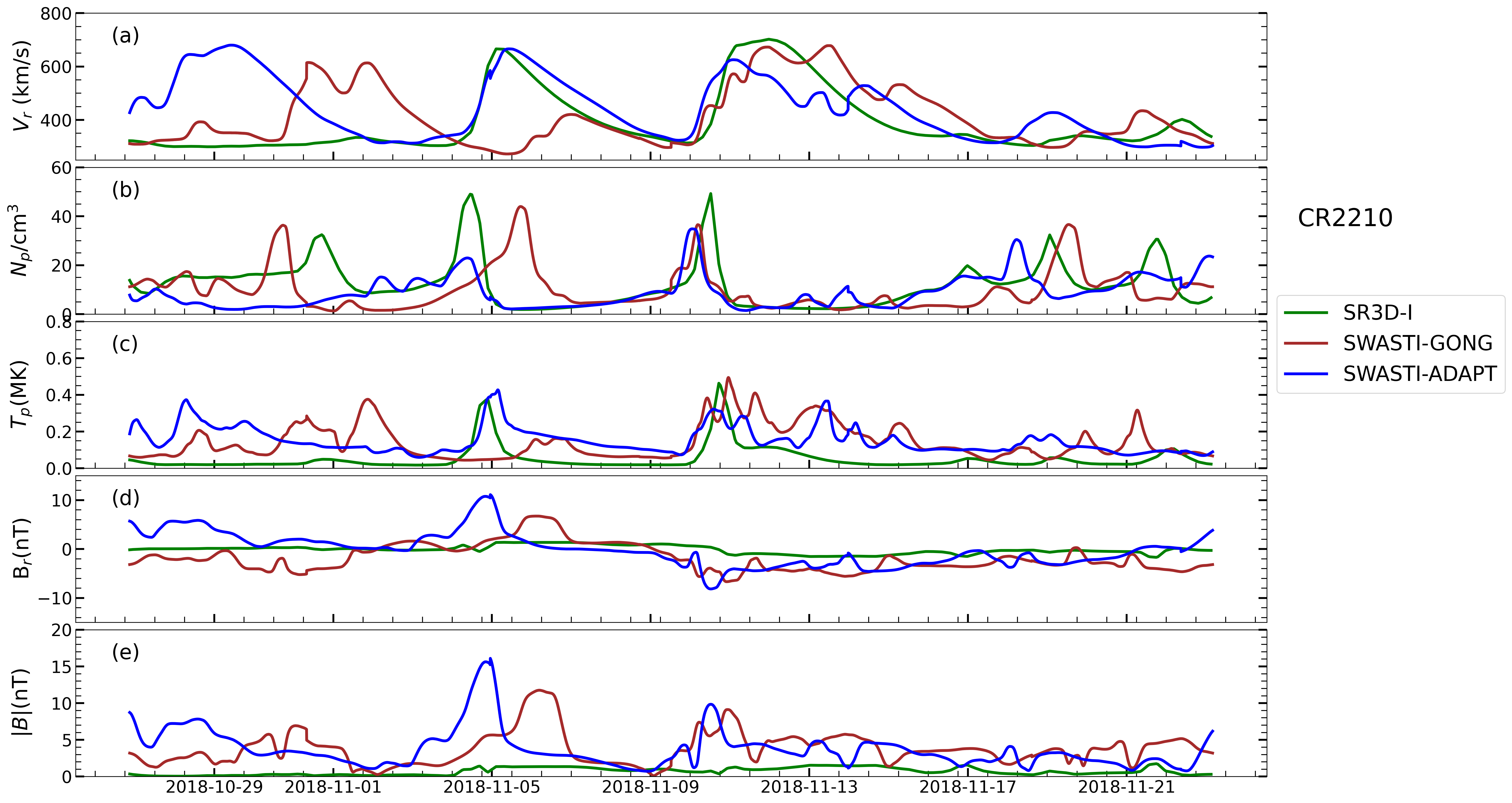}
\centerline{\bf   
      \hspace{0.49\textwidth}  \color{black}{(b)}
         \hfill}
\includegraphics[width=0.8\textwidth]
{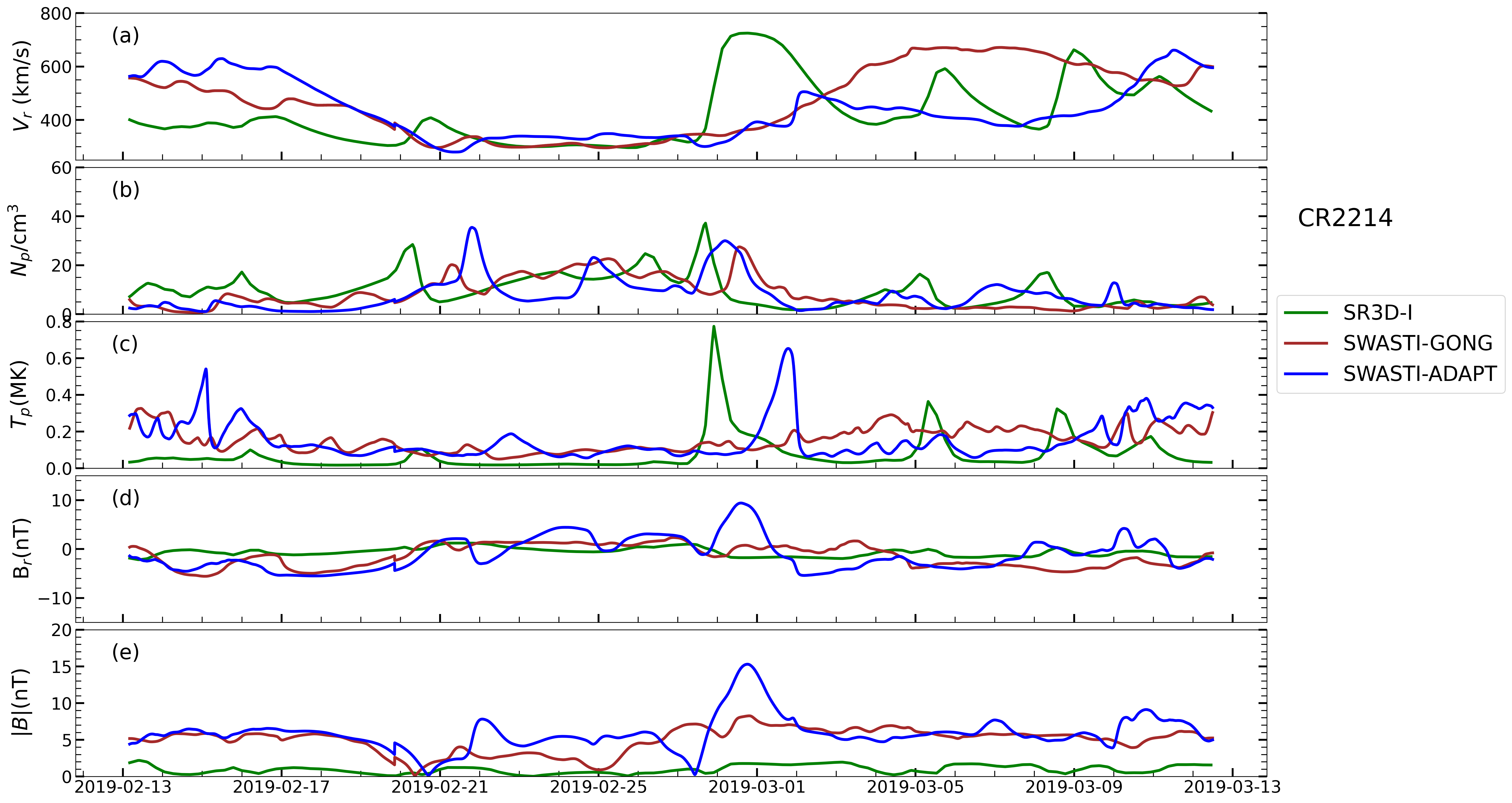}
\centerline{\bf   
      \hspace{0.49\textwidth}  \color{black}{(c)}
         \hfill}
\includegraphics[width=0.8\textwidth]{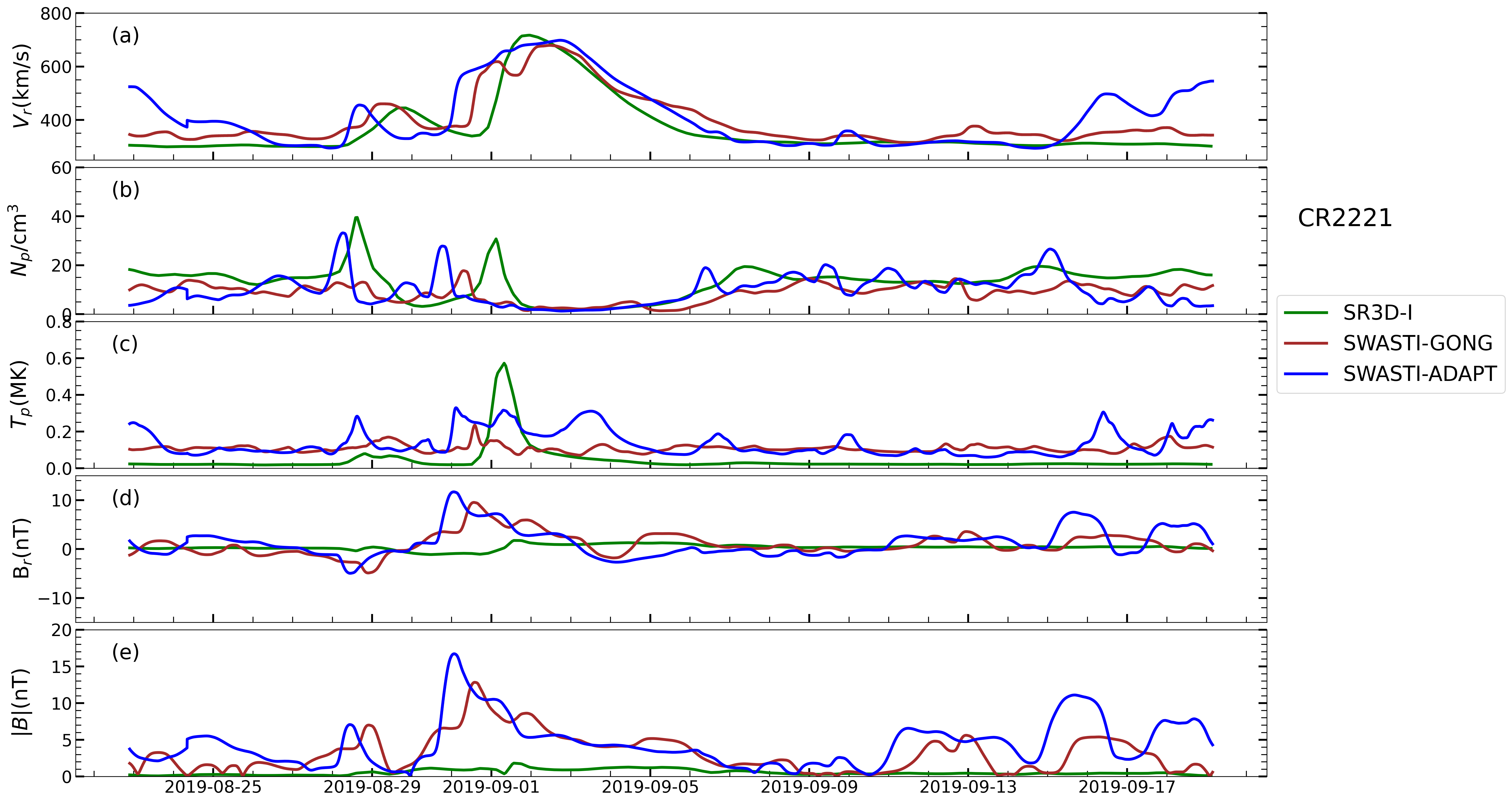}
\caption{SR3D-I (green lines) and SWASTi results using default coronal models based on GONG (brown lines) and ADAPT (blue lines) magnetograms. All plots of CR2210 (a), CR2214 (b), and CR2221 (c) are of 12-hour averaged measurements at L1.}
\label{fig:SWASTi_plots}
\end{figure}

Figure \ref{fig:SWASTi_plots} illustrates the time series graphs depicting solar wind characteristics at 1 AU, utilizing different coronal models across three CR periods: CR2210, CR2214, and CR2221. We compared three distinct solutions - SR3D-I, SWASTi-GONG, and SWASTi-ADAPT at L1. These solutions differ primarily in their coronal domains, which serve as the initial boundaries for the heliospheric domain. As indicated by their names, the first two solutions derive from GONG and ADAPT magnetograms, respectively, utilizing the modified WSA relation to estimate solar wind speed at the boundary. Conversely, SR3D-I employs a physics-based MAS coronal solution grounded on HMI magnetogram data to establish the initial boundary for the heliospheric domain. We note that the three solutions significantly differ in the three panels of Figure \ref{fig:SWASTi_plots}. Specifically, the semi-empirical (SWASTi-GONG and SWASTi-ADAPT) obtain higher solar wind speeds than the physics-based initial conditions of SR3D-I, which is similar to CR2210 (a) and CR2214 (b). However, for CR2221, all approaches demonstrated comparable results for the solar wind speed profiles.
Regarding proton number density analysis, SR3D-I obtains higher density peaks than SWASTi-GONG and SWASTi-ADAPT for the three CRs. Conversely, the temperature profiles revealed that SWASTi-ADAPT and SRD3D-I estimate higher temperatures than SWASTi-GONG, as shown in the three panels. For the radial magnetic field and the magnetic field magnitude, it is discernible that SWASTi-GONG and SWASTi-ADAPT obtain larger magnitudes than SR3D-I for the three CRs.

\FloatBarrier

While showing similar behavior between the SWASTi and SRD3D-I results in the level of matching number density, temperature, and magnetic field for all CR periods, there are differences between them in the setups, particularly in the location of the inner radial boundary in the heliospheric domain, the extent of boundaries in radial and latitudinal directions, grid resolution, and the value of the polytropic index. A significant distinction was noted in the relative flatness of the profiles for the magnetic field obtained by SR3D-I compared to those from SWASTi. This difference might be due to the lower numerical grid resolution in the SR3D-I solution and the underestimation of the magnetic field magnitude in the boundary conditions provided by CORHEL.

\section{Conclusions}
\label{Conclusions}

In this paper, we have used SR3D to interpret the global structure of the heliosphere from in-situ measurements. This model uses boundary conditions generated by CORHEL which uses coronal conditions to generate steady-state solutions that extend up to 0.14 AU. In the inner heliosphere domain, the PLUTO code is employed to compute the plasma properties of SW with the MHD approximation of up to 1.1 AU in the inertial and rotating frames of reference. According to the solutions for the steady-state SW for various CRs, we found that SR3D can capture the global properties of the SW streams in the inner heliosphere, including the CIRs and the Parker Spiral magnetic field structure.

SR3D can capture global solar wind structures observed by OMNI and STA. The statistical analysis shows that SR3D produces acceptable values for the RMSE and PCC for the selected CRs. For most CRs, SR3D matched the number density better than MAS. The results for the radial magnetic field showed an underestimation of the strength; however, the three models showed reasonable values of RMSE and PCC. Overall, the statistical analysis indicates that SR3D is comparable to MAS in most CRs, which gives us the confidence to apply it to simulations of background SW solutions in the heliosphere. Importantly, in the Taylor diagrams, it is discernible that the cluster patterns show that different solutions for a particular CR are closer than the types of runs over different CRs, which implies consistency in the solutions of the three models. In addition, it is notable that PLUTO gets the solutions using a medium grid resolution running on a local workstation, which represents an advantage compared to the models that need supercomputers to exploit their full capabilities.  

In our comparative analysis of SR3D with SWASTi, we observed that both models offer comparable accuracy in their results, even though these models have the following primary differences: SWASTi employs GONG and ADAPT magnetic field maps based on the WSA model and also incorporates gravity and adopts an ideal equation of state with $\gamma=$1.5. While SR3D employs HMI magnetic field maps, the distance from the DCHB model ignores gravity and uses $\gamma=5/3$ in the ideal equation of state. Nevertheless, there is room for SR3D to enhance the precision in computing the magnetic field magnitude. Our study indicates that physics-based initial boundary conditions might yield accurate results. However, semi-empirical boundary conditions can offer consistent results during specific periods. It is crucial to highlight that the computational time required for solving physics-based coronal domains is more significant than the semi-empirical approach, even at a considerably lower resolution. Therefore, the ultimate choice depends on the user's preferences and requirements. A sensible strategy might be to cultivate an ensemble of solutions, incorporating both approaches, to optimize the potential for accurate solar wind property forecasting. Furthermore, a relevant potential improvement of our models could be creating seamless links with different coronal models to integrate a modular code capable of running WSA, DCHB, or fully-time-dependent output from coronal models. This time-dependent output could propagate, for example, the CME directly from one region to another.   

Additionally, SR3D describes the global structure of the SW streams observed near the Earth environment, i.e., at about 1 AU, and also features observed by STA. Hence, it is a helpful model for interpreting the global structure of the heliosphere from background SW conditions, which include CIRs and SIRs. Our medium- and long-term plans aim to share SR3D as a community-developed open-source model, allowing us to receive constant feedback from, for example, the international heliospheric community dedicated to modeling. Also, given the capabilities, SR3D could be part of CCMC as a heliospheric model in the future. Finally, we envisage SR3D as a tool for scientific understanding and forecasting of ICMEs in the inner heliosphere.

\begin{acknowledgements}
We appreciate the constructive comments and suggestions of the two anonymous referees that helped to increase the quality of the results presented in this paper. JJGA acknowledges the "Programa de Apoyo a Proyectos de Investigación e Innovación Tecnológica" (PAPIIT) IN115423 project and "Consejo Nacional de Humanidades Ciencias y Tecnologías (CONAHCYT)" 319216 project "Modalidad: Paradigmas y Controversias de la Ciencia 2022," for the financial support of this work. PR and MBN gratefully acknowledge support from NASA (80NSSC20K0695, 80NSSC20K1285, 80NSSC23K0258, and the Parker Solar Probe WISPR contract NNG11EK11I to NRL (under subcontract N00173-19-C-2003 to PSI).
We thank Goddard Space Flight Center's Space Physics Data Facility (\href{https://omniweb.gsfc.nasa.gov/}{https://omniweb.gsfc.nasa.gov/}) for providing the solar wind data. In addition, we produce the synoptic maps of the radial magnetic field and the time series of the comparisons between models and in-situ measurements with the PsiPy package (\href{https://psipy.readthedocs.io/en/stable/}{https://psipy.readthedocs.io/en/stable/}); while we generate Figure \ref{CR2210_3D} using VisIt (\href{https://visit-dav.github.io/visit-website/index.html}{https://visit-dav.github.io/visit-website/index.html}) and Figure \ref{Selected_CRs} with pyPLUTO (\href{https://github.com/coolastro/pyPLUTO}{https://github.com/coolastro/pyPLUTO}) packages. The authors are also grateful to the developers of the PLUTO software, which provides a general and sophisticated interface for the numerical solution of mixed hyperbolic/parabolic systems of partial differential Equations (conservation laws) targeting high Mach number flows in astrophysical fluid dynamics.  
\end{acknowledgements}


\bibliography{jswsc}

\Online

\begin{appendix} 
\section{Rotating vs inertial frame analysis}
\label{Appendix_A}

The differences briefly mentioned in section \ref{Results} are related to the results in the time series comparisons, showing that SR3D-R and SR3D-I are sharper than MAS in regions with higher variations in density, which is a consequence of the less diffusive numerical algorithms employed by PLUTO compared to MAS. Moreover, we highlight two primary characteristics in analyzing the simulation of solar wind in both inertial and rotating frames: the diffusion phenomenon and the presence of a phase shift. Notably, the outcomes within the rotating frame exhibited a higher diffusion level when contrasted with those in the inertial frame. Additionally, the time series plots demonstrated phase shift between rotating and inertial frames.

The peaks in solar wind properties exhibited greater sharpness in the inertial frame. This discrepancy in behavior might stem from numerical diffusion resulting from the rotation of the solar wind within the rotating frame. In the inertial frame, solar wind properties are adjusted with the appropriate angle at each time step, independently of the simulation's numerical calculations. Conversely, the rotating frame produces the solar wind rotation due to the source terms defined by the Centrifugal and Coriolis forces. These source terms introduce extra diffusion, in contrast to rotating the inner boundary conditions in the inertial frame. The impact of this added diffusion becomes increasingly noticeable with greater radial distance from the simulation's inner boundary. As a result, at a distance of 1 AU, discernible shifts in the profile of solar wind properties are observed, while these differences remain minimal closer to the simulation's inner boundary.

\begin{figure}
   \centering
   \includegraphics[width = \textwidth]{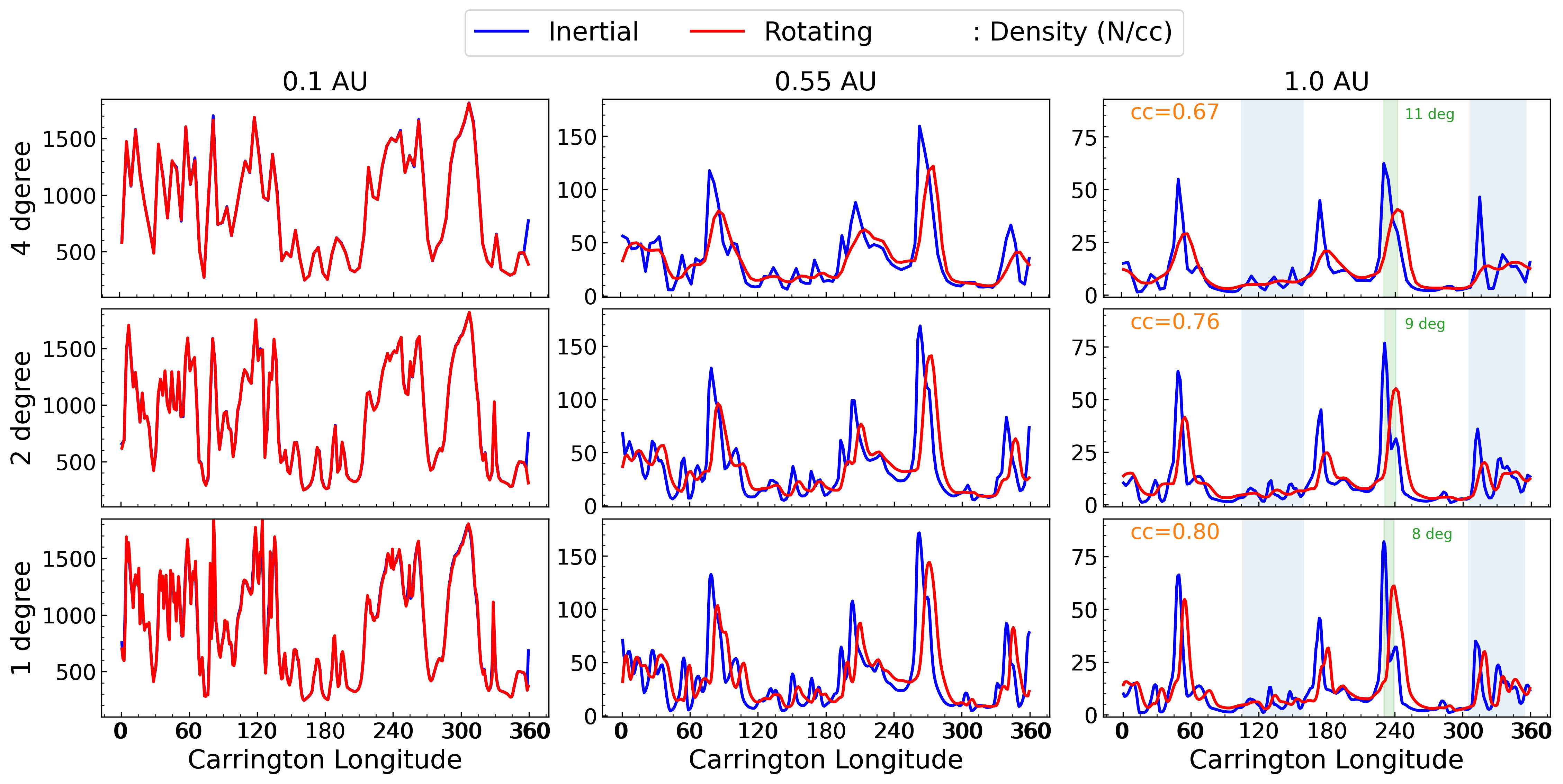}
   \caption{The picture demonstrates the differences between the solar wind solutions of the inertial and rotating frames. The differences become evident at larger radial distances.}
   \label{fig:inertial_vs_rot}
\end{figure}

Figure \ref{fig:inertial_vs_rot} illustrates the profile of solar wind density for different resolutions (rows) at varying radial distances (columns) in the equatorial plane. At 0.10 AU, the inner boundary, there is no difference between the inertial and rotating plots. However, the differences are apparent at 0.55 and 1.0 AU. The peaks in the inertial frame results are sharp, and the profile exhibits smaller patterns than the rotating frame results. Furthermore, Figure \ref{fig:inertial_vs_rot} also illustrates a phase shift between the two profiles. When we initially assign $V_{\phi}=0$ at the internal boundary $R_{b}$ in the rotating frame, we achieve solutions with a noticeable phase shift relative to those in the inertial frame. This condition was mitigated by setting $V_{\phi}=-r\Omega$ in the rotating frame, bringing the solutions closer to those in the inertial frame with only a minor phase shift. The latter means that if we employ a purely radial flow at the inner boundary in the inertial frame, it is necessary to define $V_{\phi} = -r\Omega$ in the rotating frame to make the two models consistent. This aspect of the results has yet to be documented in global heliospheric modeling and, we believe, merits further detailed investigation.

We simulated the same solar wind condition for different resolutions to explore the reason behind the abovementioned differences further. In Figure \ref{fig:inertial_vs_rot}, the discrepancies between the two rotating frames diminished as the resolution increased. In the rotating frame, the density peaks became sharper. For the simulation at the 4-degree resolution, the density profile appeared more diffuse than that in the inertial frame, especially in regions of low density (blue-shaded regions). At 2-degree resolution, this disparity slightly lessened, and the features became more pronounced. The plots at 1-degree resolution are markedly more similar, albeit with some differences in the phi-direction shift.

The highest degree of shift is also highlighted in Figure \ref{fig:inertial_vs_rot}, showing 11, 9, and 8 degrees for 4, 2, and 1-degree resolutions, respectively. To further quantify the correlation between the solutions in both frames, we computed the Pearson correlation coefficient (cc) for both. They were calculated after adjusting the rotating frame by its maximum shift value to lend more significance to the cc values. We observed that the cc value increased from 0.67 to 0.76 to 0.8 with an increase in resolution. This gradual convergence of density plots in rotating and inertial frames strongly suggests that numerical diffusion is the primary factor behind the observed discrepancies. With further increased resolutions, the alignment between these plots is expected to improve even more.

The computational time required to model the same solar wind conditions in each frame differs significantly. Rotating frames demands considerably more time than inertial frames. For instance, in a simulation with 1-degree resolution, the inertial frame required approximately 4 hours, whereas the rotating frame took about 44 hours to complete on a 48-core CPU. Given that higher resolution results in the rotating frame increasingly resembling those from the inertial frame, opting for the inertial frame in medium-resolution simulations seems a prudent choice.

\end{appendix}

\begin{appendix} 
\section{Full statistical results of the comparisons of models with in-situ measurements}
\label{Appendix_B}

This appendix shows the full statistical results obtained in comparing SR3D-I, SR3D-R, and MAS with OMNI and STA 12-hour averaged data for all the CRs considered in this paper. Table \ref{tab:Stat_models_vs_OMNI} contains the statistical results represented by RMSE/PCC between SR3D-R, SR3D-I, and MAS with OMNI's 12- hour averaged data for the radial velocity $V_{r}$ in km s$^{-1}$, proton number density $N_{p}$ in cm$^{-3}$, proton temperature $T_{p}$ in MK, radial magnetic field $B_{r}$ in nT and magnetic field magnitude $|B|$ in nT. Table \ref{tab:Stat_models_vs_STEREO-A} contains the STA 12-hour averaged data comparisons statistical results.

\begin{center}
\begin{longtable}{|c|c|c|c|c|c|c|}
\caption{\label{tab:Stat_models_vs_OMNI} Statistical Results of RMSE/PCC for the comparison of SR3D-R, SR3D-I and MAS models with the OMNI 12-hours averaged data}
\\
\hline \multicolumn{1}{|c|}{\textbf{CR}} & \multicolumn{1}{c|}{\textbf{Model}} &  \multicolumn{1}{c|}{\textbf{$V_{r}$(km s$^{-1}$)}}  & \multicolumn{1}{c|}{\textbf{$N_{p}$}(cm$^{-3}$)}  &  \multicolumn{1}{c|}{\textbf{$T_{p}$(MK)}}  &  \multicolumn{1}{c|}{\textbf{$B_{r}$(nT)}}  &  \multicolumn{1}{c|}{\textbf{$|B|$(nT)}}   \\
\hline\endfirsthead
\multicolumn{7}{c}%
{{\bfseries \tablename\ \thetable{} -- continued from previous page}} \\
\hline \multicolumn{1}{|c|}{\textbf{CR}} & \multicolumn{1}{c|}{\textbf{Model}} &  \multicolumn{1}{c|}{\textbf{$V_{r}$(km s$^{-1}$)}}  & \multicolumn{1}{c|}{\textbf{$N_{p}$(cm$^{-3}$)}}  &  \multicolumn{1}{c|}{\textbf{$T_{p}$(MK)}}  &  \multicolumn{1}{c|}{\textbf{$B_{r}$(nT)}}  &  \multicolumn{1}{c|}{\textbf{$|B|$(nT)}}   \\ \hline 
\endhead

\hline \multicolumn{7}{|r|}{{Continued on next page}} \\ \hline
\endfoot

\hline 
\endlastfoot
2190  & SR3D-R & 87.97/0.74 & 7.80/0.46 & 0.052/0.74 & 2.35/0.56 & 3.41/-0.08 \\
 &   SR3D-I & 87.05/0.72 & 9.26/0.32 & 0.066/0.60 & 2.38/0.52  & 3.25/0.07 \\
 &   MAS & 85.80/0.74 & 8.52/0.36  & 0.052/0.68 & 2.34/0.60  & 3.20/0.14 \\
2199  & SR3D-R & 83.70/0.35 & 7.23/0.15 & 0.086/0.19 & 1.87/0.52 & 2.35/0.03 \\
 &   SR3D-I & 87.05/0.36 & 8.69/0.11 & 0.086/0.02 & 1.98/0.42 & 2.86/0.06 \\
 &   MAS & 86.91/0.37 & 8.64/0.09 & 0.063/0.22 & 1.96/0.44 & 2.19/0.07 \\
2202  & SR3D-R & 102.36/0.50 & 7.53/0.47 & 0.096/0.29 & 1.79/0.65 & 2.53/0.17 \\
 &   SR3D-I & 113.44/0.45 & 8.39/0.43 & 0.083/0.21 & 1.82/0.52 & 2.82/0.27 \\
 &   MAS & 109.65/0.51 & 8.32/0.39 & 0.064/0.47 & 1.78/0.57 & 2.27/0.28 \\
2203  & SR3D-R & 108.64/0.64 & 9.35/0.10 & 0.102/0.42 & 2.06/0.51 & 2.86/0.14 \\
 &   SR3D-I & 110.67/0.63 & 10.38/0.02 & 0.097/0.35 & 2.17/0.38 & 2.87/0.08 \\
 &   MAS & 111.45/0.65 & 9.75/0.04 & 0.074/0.49 & 2.10/0.47 & 2.65/0.11 \\
2205  & SR3D-R & 105.43/0.33 & 8.92/0.21 & 0.063/0.25 & 2.29/0.55 & 2.29/0.18 \\
 &   SR3D-I & 106.19/0.31 & 9.71/0.19 & 0.065/0.26 & 2.29/0.53 & 2.29/0.38 \\
 &   MAS & 108.61/0.33 & 9.41/0.24 & 0.056/0.26 & 1.92/0.61 & 2.73/0.35 \\
 2208  & SR3D-R & 96.67/0.54 & 8.02/0.38 & 0.072/0.60 & 2.08/0.38 & 2.16/0.02 \\
 &   SR3D-I & 101.33/0.51 & 9.11/0.31 & 0.062/0.61 & 1.97/0.49 & 2.09/0.45 \\
 &   MAS & 110.88/0.50 & 8.59/0.34 & 0.059/0.42 & 1.87/0.58 & 2.35/0.42 \\
 2209 & SR3D-R & 80.70/0.76 & 9.25/0.30 & 0.07/0.56 & 1.68/0.46 & 1.99/-0.001 \\
 &   SR3D-I & 84.04/0.72 & 11.05/0.15  & 0.07/0.45 & 1.65/0.49 & 2.07/0.41 \\
 &   MAS & 79.93/0.76 & 9.66/0.22 & 0.05/0.59 & 1.49/0.66 & 2.01/0.51 \\
 2210  & SR3D-R & 72.54/0.75 & 7.79/0.52 & 0.079/0.51 & 1.66/0.61 & 2.28/0.19 \\
 &   SR3D-I & 79.80/0.74 & 9.32/0.45 & 0.063/0.48 & 1.79/0.47 & 2.61/0.25 \\
 &   MAS & 100.72/0.66 & 8.92/0.42 & 0.049/0.65 & 1.69/0.58 & 2.44/0.30 \\
 2211  & SR3D-R & 101.51/0.47 & 6.92/0.26 & 0.09/0.14 & 1.55/0.79 & 2.63/0.01 \\
 &   SR3D-I & 108.53/0.45 & 8.09/0.43 & 0.07/0.25 & 1.72/0.61 & 2.94/0.22 \\
 &   MAS & 106.28/0.44 & 8.35/0.32 & 0.06/0.45 & 1.71/0.65 & 2.41/0.25 \\
 2214  & SR3D-R & 103.09/0.46 & 6.40/0.42 & 0.102/0.46 & 1.81/0.47 & 3.06/0.36 \\
 &   SR3D-I & 113.65/0.44 & 7.16/0.41 & 0.085/0.41 & 1.81/0.51 & 3.32/0.26 \\
 &   MAS & 124.71/0.44 & 7.10/0.46 & 0.068/0.53 & 1.75/0.57 & 2.26/0.31 \\
 2215  & SR3D-R & 141.36/0.54 & 5.92/0.55 & 0.105/0.15 & 1.81/0.65 & 3.08/0.35 \\
 &   SR3D-I & 146.77/0.54 & 6.93/0.48 & 0.082/0.20 & 1.91/0.52 & 3.25/0.24 \\
 &   MAS & 155.94/0.53 & 6.89/0.45 & 0.085/0.39 & 1.89/0.52 & 2.71/0.26 \\
 2221  & SR3D-R & 124.29/0.73 & 10.05/0.23 & 0.083/0.65 & 2.07/0.27 & 2.00/0.25 \\
 &   SR3D-I & 125.37/0.71 & 10.22/0.13 & 0.084/0.57 & 2.07/0.43 & 2.03/0.33 \\
 &   MAS & 124.06/0.74 & 9.84/0.17 & 0.061/0.67 & 1.97/0.49 & 2.26/0.25 \\
\hline
\end{longtable}
\end{center}

\begin{center}
\begin{longtable}{|c|c|c|c|c|c|c|}
\caption{\label{tab:Stat_models_vs_STEREO-A} Statistical Results of Comparison of RMSE/PCC for SR3D-R, SR3D-I and MAS models with the STA 12-hours averaged data}
\\
\hline \multicolumn{1}{|c|}{\textbf{CR}} & \multicolumn{1}{c|}{\textbf{Model}} &  \multicolumn{1}{c|}{\textbf{$V_{r}$(km s$^{-1}$)}}  & \multicolumn{1}{c|}{\textbf{$N_{p}$}(cm$^{-3}$)}  &  \multicolumn{1}{c|}{\textbf{$T_{p}$(MK)}}  &  \multicolumn{1}{c|}{\textbf{$B_{r}$(nT)}}  &  \multicolumn{1}{c|}{\textbf{$|B|$(nT)}}   \\
\hline\endfirsthead

\multicolumn{5}{c}%
{{\bfseries \tablename\ \thetable{} -- continued from previous page}} \\
\hline \multicolumn{1}{|c|}{\textbf{CR}} & \multicolumn{1}{c|}{\textbf{Model}} &  \multicolumn{1}{c|}{\textbf{$V_{r}$(km s$^{-1}$)}}  & \multicolumn{1}{c|}{\textbf{$N_{p}$(cm$^{-3}$)}}  &  \multicolumn{1}{c|}{\textbf{$T_{p}$(MK)}}  &  \multicolumn{1}{c|}{\textbf{$B_{r}$(nT)}}  &  \multicolumn{1}{c|}{\textbf{$|B|$(nT)}}   \\ \hline 
\endhead

\hline \multicolumn{7}{|r|}{{Continued on next page}} \\ \hline
\endfoot

\hline \hline
\endlastfoot
2190  & SR3D-R & 97.87/0.78 & 7.73/0.49 & 0.076/0.64 & 1.66/0.73 & 3.01/0.05 \\
 &   SR3D-I & 94.67/0.76 & 8.93/0.41 & 0.093/0.47 & 1.85/0.58 & 2.85/0.18 \\
 &   MAS & 94.42/0.76 & 8.36/0.39 & 0.074/0.59 & 1.78/0.65 & 2.76/0.32 \\
 2199  & SR3D-R & 168.69/0.07 & 6.88/0.45 & 0.107/0.35 & 1.60/0.61 & 2.17/0.09 \\
 &   SR3D-I & 172.39/0.08 & 7.54/0.47 & 0.112/0.32 & 1.73/0.49 & 2.23/-0.03 \\
 &  MAS & 165.44/0.13 & 6.72/0.46 & 0.086/0.39 & 1.71/0.51 & 2.19/-0.04 \\
2202  & SR3D-R & 140.97/0.21 & 8.07/0.15 & 0.136/-0.15 & 1.71/0.38 & 2.09/0.11 \\
 &   SR3D-I & 145.01/0.22 & 8.86/0.17 & 0.126/-0.11 & 1.68/0.40 & 2.15/0.06 \\
 &   MAS & 148.01/0.18 & 9.27/0.14 & 0.104/0.01 & 1.66/0.41 & 2.11/0.06 \\
 2203  & SR3D-R & 156.38/0.11 & 8.47/0.52 & 0.101/0.40 & 1.48/0.32 & 2.54/0.17 \\
 &   SR3D-I & 159.27/0.13 & 9.47/0.56 & 0.094/0.35 & 1.54/0.32 & 2.22/0.48 \\
 &   MAS & 163.22/0.13 & 9.29/0.53 & 0.082/0.31 & 1.55/0.28 & 2.33/0.37 \\
 2205  & SR3D-R & 104.94/0.79 & 7.35/0.62 & 0.078/0.64 & 1.70/0.53  & 2.69/-0.07 \\
 &   SR3D-I & 105.32/0.78 & 8.12/0.50 & 0.078/0.63 & 1.70/0.37  &  2.69/-0.03 \\
 &   MAS & 110.89/0.76 & 8.18/0.52 & 0.056/0.73 & 1.53/0.41 & 1.97/0.05 \\
 2208  & SR3D-R & 187.30/0.52  & 10.67/0.21 & 0.108/0.51 & 1.53/0.41  & 2.28/-0.19 \\
 &   SR3D-I & 186.99/0.53 & 12.06/0.19 & 0.096/0.44 & 1.72/0.24 &  2.46/-0.18\\
 &   MAS & 177.98/0.52 & 10.81/0.26 & 0.065/0.59 & 1.61/0.31 & 2.27/-0.16  \\
 2209  & SR3D-R & 308.00/-0.61  & 11.03/0.46  & 0.137/-0.34  & 2.03/0.66 & 3.12/0.08 \\
 &   SR3D-I & 317.05/-0.59 & 12.17/0.36 & 0.104/-0.30  & 2.06/0.61  & 2.94/0.34\\
 &   MAS & 322.66/-0.63 & 11.80/0.33 & 0.101/-0.40 & 2.05/0.65 & 2.89/0.39 \\
2210  & SR3D-R & 88.86/0.64 & 10.14/0.55 & 0.069/0.61 & 1.88/0.41 & 2.31/0.10 \\
 &   SR3D-I & 88.60/0.66 & 10.24/0.57 & 0.068/0.57 & 1.72/0.56 & 2.09/0.48 \\
 &   MAS & 85.13/0.69 & 9.07/0.55 & 0.052/0.65 & 1.69/0.60 & 2.07/0.49 \\
 2211  & SR3D-R & 94.87/0.38 & 11.62/0.23 & 0.055/0.18 & 1.85/0.53 & 2.53/0.35 \\
 &   SR3D-I & 95.82/0.37 & 11.25/0.21 & 0.056/0.17 & 1.77/0.58 & 2.29/0.52 \\
 &   MAS & 98.72/0.35 & 10.19/0.17 & 0.043/0.22 & 1.77/0.56 & 2.22/0.44 \\
  2214  & SR3D-R & 114.45/0.05 & 9.85/0.32 & 0.083/-0.14 &  1.94/0.60 & 2.99/0.10 \\
 &   SR3D-I & 116.74/0.03  & 10.06/0.20 & 0.080/-0.13 & 1.95/0.60  & 2.76/0.24 \\
 &   MAS & 110.81/0.04 & 9.57/0.18 & 0.076/-0.17 & 1.92/0.65 & 2.79/0.10 \\
  2215  & SR3D-R & 123.26/0.14 & 7.12/0.46  & 0.074/0.14  & 1.66/0.53 & 2.85/-0.07  \\
 &   SR3D-I & 130.65/0.12 & 7.68/0.49 & 0.074/0.13 & 1.57/0.67 & 2.69/0.06 \\
 &   MAS & 130.76/0.12  & 7.72/0.46 & 0.068/0.07 & 1.55/0.67  & 2.68/-0.02 \\
  2221  & SR3D-R & 83.39/0.72  & 8.47/0.62 & 0.051/0.76 & 1.93/0.48 & 2.95/-0.17 \\
 &   SR3D-I & 83.52/0.71  & 9.45/0.58  & 0.055/0.70 & 1.98/0.43 &  2.85/0.24 \\
 &   MAS & 80.22/0.75 & 8.68/0.66 & 0.036/0.84 & 1.85/0.65 & 2.87/0.22 \\
 \hline
\end{longtable}
\end{center}
\end{appendix}
   

\end{document}